\documentclass{aa}  

\usepackage{graphicx}
\usepackage{enumerate}
\usepackage{natbib}
\usepackage{mathrsfs}
\usepackage[breaklinks=true]{hyperref} %% to avoid \citeads line fills
\bibpunct{(}{)}{;}{a}{}{,} %% natbib format for A&A and ApJ
\bibpunct{(}{)}{;}{a}{}{,}
\usepackage{graphicx}
\usepackage{txfonts}
\begin{document}

   \title{Open cluster Dolidze~25: Stellar parameters and the metallicity in the Galactic Anticentre\thanks{Based on observations made with the Nordic Optical Telescope, the Mercator Telescope, and the telescopes of the Isaac Newton Group.}}

    \author{I. Negueruela
          \inst{1}
          \and
          S.~Sim\'on-D\'{\i}az\inst{2,3}
\and J.~Lorenzo\inst{1} \and N.~Castro\inst{4} \and A.~Herrero\inst{2,3}
          }

   \institute{Departamento de F\'{\i}sica, Ingenier\'{\i}a de Sistemas y
  Teor\'{\i}a de la Se\~{n}al, Escuela Polit\'ecnica Superior, Universidad de Alicante, Carretera de San Vicente del Raspeig s/n,  E03690, San Vicente del Raspeig, Alicante, Spain\\
\email{ignacio.negueruela@ua.es}
         \and Instituto de Astrof\'{\i}sica de Canarias,  V\'{\i}a L\'actea s/n, E38200, La Laguna, Tenerife, Spain 
\and
Departamento de Astrof\'{\i}sica, Universidad de La Laguna, Facultad de F\'{\i}sica y Matem\'aticas, Universidad de La Laguna, Avda. Astrof\'{\i}sico Francisco S\'anchez, s/n, E38205, La Laguna, Tenerife, Spain
\and
Argelander Institut f\"ur Astronomie, Auf den H\"ugel 71, Bonn, 53121, Germany 
}
             
  \titlerunning{Abundances and parameters in Dolidze~25}          

   \date{}

% \abstract{}{}{}{}{} 
% 5 {} token are mandatory
 
  \abstract
   {The young open cluster Dolidze~25, in the direction of the Galactic Anticentre, has been attributed a very low metallicity, with typical abundances between $-0.5$ and $-0.7$ dex below solar.}
   {We intend to derive accurate cluster parameters and accurate stellar abundances for some of its members.}
   {We have obtained a large sample of intermediate- and high-resolution spectra for stars in and around Dolidze~25. We used the \textsc{fastwind} code to generate stellar atmosphere models to fit the observed spectra. We derive stellar parameters for a large number of OB stars in the area, and abundances of oxygen and silicon for a number of stars with spectral types around B0.}
   {We measure low abundances in stars of Dolidze~25. For the three stars with spectral types around B0, we find 0.3~dex (Si) and 0.5~dex (O) below the values typical in the solar neighbourhood. These values, even though not as low as those given previously, confirm Dolidze~25 and the surrounding \ion{H}{ii} region Sh2-284 as the most metal-poor star-forming environment known in the Milky Way. We derive a distance $4.5\pm0.3\:$kpc to the cluster ($r_{\textrm{G}}\approx12.3\:$kpc). The cluster cannot be older than $\sim3$~Myr, and likely is not much younger. One star in its immediate vicinity, sharing the same distance, has Si and O abundances at most 0.15~dex below solar.}
   {The low abundances measured in Dolidze~25 are compatible with currently accepted values for the slope of the Galactic metallicity gradient, if we take into account that variations of at least $\pm0.15$~dex are observed at a given radius. The area traditionally identified as Dolidze~25 is only a small part of a much larger star-forming region that comprises the whole dust shell associated with Sh2-284 and very likely several other smaller \ion{H}{ii} regions in its vicinity.}


   \maketitle
%
%________________________________________________________________

\section{Introduction}

The young open cluster Dolidze~25 (Do~25 = C0642+0.03 = OCL-537) is located near the centre of the large \ion{H}{ii} region Sh2-284 (Galactic coordinates: $l=211\fdg9$, $b=-01\fdg3$), and assumed to be its ionisation source. With distance estimates to Do~25 ranging from $\sim$$4$ to $\sim$$6$ kpc, it is believed to be one of the massive star formation regions at the largest galactocentric radii, and an excellent testbed for massive star formation in the Galactic Anticentre. Using photoelectric photometry of the brightest members, \citet{moffat75} determined that the cluster contained O-type stars and was located at a distance $d=5.25$~kpc. \citet{lennon90} used Str\"omgren photometry to select spectroscopic targets and took intermediate-resolution spectra of two O-type and one B0 star in the cluster. From atmosphere model fitting, they determined surprisingly low abundances for some elements, implying an overall metallicity around one-sixth of solar \citep{lennon90}.

Later, \citet{turbide} obtained CCD photometry of the cluster core. From a fit to a sequence of 12 likely members, they estimated an average colour excess $\left< E(B-V)\right> =0.83\pm0.08$ and an age of 6~Myr. For solar metallicity, this would imply a distance of $d=5.5\pm0.4$~kpc, while for the lower metallicity expected at this Galactocentric radius, $d=5.0\pm0.4$~kpc. Making use of more extensive photometry, \citet{delgado10} determined $E(B-V)=0.78\pm0.02$ and an age of 3.2~Myr. Isochrone fits, using the value of metallicity determined by  \citet{lennon90}, place the cluster at $d=3.6$~kpc, while the use of solar neighbourhood abundances would place the cluster at a distance approaching 5~kpc, and thus compatible with that of \citet{turbide}. 

Even though all photometric studies have been limited to this central concentration, {\it Spitzer}/IRAC observations \citep{puga09} revealed the complexity of the region, showing evidence of widespread star formation over the whole extent of Sh2-284. They detected dust emission extending not only across the observed \ion{H}{ii} region (which is almost circular with a diameter $\sim20\arcmin$), but also to larger distances (over $30\arcmin$). The dust emission surrounds the ionised gas, even though it is less prominent to the south-west, a fact that \citet{puga09} attribute to a density gradient within the molecular cloud. In addition to this very large bubble, they found two much smaller, almost circular bubbles close to the edges of the \ion{H}{ii} region. From an analysis of point-like sources over the whole field and cometary globules detected at the rim of the main bubble, \citet{puga09} conclude that star formation is currently taking place in the area, that the age spread is small, and that triggered star formation is likely taking place at the bright rim of the \ion{H}{ii} region. Later, \citet{cusano11} took spectra of several low-mass pre-main-sequence (PMS) candidates in the region, confirming their nature and concluding that triggered star formation was taking place. Recently, \citet{kalari15} used spectra of low-mass PMS stars to conclude that accretion rates are not significantly different from those observed in similar stars of solar metallicity.

With a physical size comparable to that of the Rosette Nebula, but at a Galactocentric distance of at least 12~kpc, Sh2-284 is an excellent laboratory to study star formation in the outer Galaxy. Our current understanding, however, still presents many gaps. If the cluster contains only two O9\,V stars, as assumed by \citet{lennon90}, why is the \ion{H}{ii} region so large? In contrast, the ionising cluster ot the Rosette Nebula, NGC~2244, contains seven O-type stars, including two early ones \citep[e.g.][]{martins12}. Moreover, why should a Milky Way cluster have a metallicity comparable to that of the Small Magellanic Cloud? Even though the outer regions of the Milky Way are known to have lower metallicity than the solar neighbourhood, values typically accepted for the slope of the Galactic metallicity gradient \citep[e.g.][]{rolleston00, esteban13} support abundances $\sim 0.3$~dex below solar for distances $r_{\textrm{G}}=12$\,-- \,13~kpc\footnote{Throughout the paper, we will stick to the convention and refer to the abundances as ``solar'' value. However, both for the analysis presented here and for the previous work of \citet{lennon90} or \citet{rolleston00}, the reference values for abundances are those obtained from B-type stars in the solar neighbourhood. This is particularly relevant because abundances for those stars with poor spectra are derived differentially with respect to stars in the solar neighbourhood with accurate values. In the case of this work, solar abundances are understood as those derived for B-type stars in \citet{nieva12}.}. The comparative analysis of a large sample of stars with spectral types around B0 located towards the Anticentre \citep{rolleston00} only found one other cluster with similarly low abundances, the cluster associated with the \ion{H}{ii} region Sh2-289, which lies at a much higher distance \citep{moffat79,rolleston00}.

In this paper, we address these problems with a comprehensive spectroscopic survey of early-type stars in the area of Sh2-284 and its immediate surroundings, including high-resolution spectroscopy for abundance determination. We intend to answer the following questions: 
\begin{itemize}
\item Which is the distance to Do~25?
\item What is the metallicity of Do~25? Is it consistent with the slope of the Galactic metallicity gradient at this distance?
\item Is there active star formation today in Sh2-284? For how long has it been going on? 
\item What is the extent of the star-forming region?
\end{itemize}

In Section~\ref{obs}, we present our observations, whose analysis is detailed in Section~\ref{results}. We explore the vicinity of Do~25 in Sections~\ref{sec:around} and~\ref{pms}. We then derive stellar parameters for the whole sample, and abundances for some stars in Section~\ref{spctan}. Finally, we compare our results to those of previous photometric studies, and discuss our findings and their implications in Section~\ref{disc}, wrapping up with the conclusions.

 Throughout the paper, proposed members of Do~25 will be identified by their number in the WEBDA database, when available, marked with a prefix ``S''. Other stars will be identified with their standard names in SIMBAD. Spectroscopic distances are calculated when precision optical photometry is available by using the intrinsic colour calibration of \citet{fitzgerald} and the absolute magnitude calibration of \citet{turner80}, and from 2MASS data by using the intrinsic colours for O-type stars from \citet{marplez06} or B-type stars from \citet{winkler97}, combined with the \citet{turner80} calibration.

%__________________________________________________________________

\section{Observations}
\label{obs}

Targets for spectroscopy were selected by making use of available photometry in the area. The original list of possible members of \citet{moffat75} was scanned by generating colour-magnitude diagrams (CMDs) with $JHK_{{\rm S}}$ photometry from the
2MASS catalogue \citep{skru06}. Stars with colours indicative of early type, and positions in the CMDs compatible with the members confirmed in the works of \citet{turbide} and \citet{delgado10} were selected for an initial survey. In addition, we selected bright catalogued emission-line stars, as well as other stars with colours suggestive of early-type that could be associated with bright dust emission observed in the {\it Spitzer} images. Afterwards, inspection of spectra for three stars in the immediate vicinity of Sh2-284 that has been observed as part of the IACOBsweG survey \citep{ssimon15} suggested that some early-type stars in the surrounding area could be at exactly the same distance as the cluster. This led to the observation of some catalogued OB stars in the area surrounding Sh2-284 whose 2MASS magnitudes were compatible with O-type stars at the distance of the cluster.

An initial survey of the brightest catalogued cluster members was carried out with the ISIS double-beam spectrograph, mounted on the 4.2-m William Herschel Telescope (WHT) in La Palma (Spain), on the night of 2009, September 12. The blue arm of the instrument was equipped with the R300B grating and the EEV12 CCD, covering the 308\,--\,510~nm range in the unvignetted part of the detector, with a nominal dispersion of 0.86\:\AA/pixel. With a $1\farcs0$ slit, the resolution element is four pixels, and so the resolving power in the classification region is $R\sim1\,300$. The red arm was equipped with the R600R grating and  the {\it Red+} CCD, a configuration that gives a range of 120~nm, centred on 650~nm, in the unvignetted section of the CCD with a nominal dispersion of 0.5\AA/pixel. The measured resolving power is $R\approx4\,500$ 

A few more stars were observed in service mode with ISIS on the night of 2014, December 28. On this occasion, the blue arm was equipped with the R600B grating, providing a nominal dispersion of 0.45\AA/pixel over the 387\,--513~nm range. With a slit width of $1\farcs0$ slit, the measured resolving power is $R\approx2\,700$. The red arm had the same configuration as in 2009, and the grating was centred to cover the 640\,--\,760~nm range. The log of observations with the WHT is presented in Table~\ref{loglow}.

High-resolution spectra of stars in Dolidze~25 were taken with the high-resolution FIbre-fed Echelle Spectrograph (FIES) attached to the 2.56~m Nordic Optical Telescope (NOT; La Palma, Spain) during a run on 2011, January 11\,--\,13. FIES is a cross-dispersed high-resolution echelle spectrograph, mounted in a heavily insulated building separated from and adjacent to the NOT dome, with a maximum resolving power $R=67\,000$. The entire spectral range 370\,--\,730~nm is covered without gaps in a single, fixed setting \citep{telting14}. In the present study, we used the low-resolution mode with $R=25\,000$. The spectra were homogeneously reduced using the
FIEStool\footnote{http://www.not.iac.es/instruments/fies/fiestool/FIEStool.html} software in advanced mode. A complete set of bias, flat, and arc frames, obtained each night, was used to this aim.  For wavelength calibration, we used arc spectra of a ThAr lamp with a signal-to-noise ratio (S/N) typically in the $\sim100$\,--\,150 range. The FIEStool pipeline provides wavelength calibrated, blaze corrected, order merged spectra, and can also be used to normalise and correct for heliocentric velocity the final spectra.

\begin{table}
\centering
 \begin{minipage}{\columnwidth}
\caption{Log of intermediate-resolution observations taken with the WHT. Classification observations in 2009 were taken with the R300B grating, while the 2014 observations were taken with the R600B grating.\label{loglow}} 
\scalebox{0.85}{\begin{tabular}{lccc}
\hline\hline
\noalign{\smallskip}
Star & Other &Date of& Exposure\\ 
&name &observation& time (s)\\
\noalign{\smallskip}
\hline
\noalign{\smallskip}
S1 & LS VI +00 19 &2009 Sep 12&300\\
S9 & GSC 00147-01042&2009 Sep 12&350\\
S12& BD+00 1573s &2009 Sep 12&300\\
S13& GU Mon &2009 Sep 12&350\\
S15& TYC 148-2577-1 &2009 Sep 12&300\\
S17& TYC 148-2558-1 &2009 Sep 12&300\\
S22& TYC 148-503-1 &2009 Sep 12&350\\
S24& TYC 148-2254-1 &2009 Sep 12&300\\
&& 2014 Dec 28&700\\
S33& TYC 148-2536-1 &2009 Sep 12&300\\
TYC 147-2024-1& J06435739+0021058 & 2014 Dec 28&900\\
UCAC4 452-021858 & J06435751+0020508 & 2014 Dec 28& 900\\
SS 57 & TYC 147-506-1 &2009 Sep 12&400\\
&& 2014 Dec 28& 600\\
SS 62 & GSC 00148-02146& 2014 Dec 28& 750\\
$\left[ \mathrm{KW97}\right]$ 33-6 &  GSC 00148-02110 & 2014 Dec 28& 900\\
\noalign{\smallskip}
\hline
\end{tabular}}
\end{minipage}
\end{table}

Some stars in the neighbourhood of Dolidze~25 were observed in service mode using the same configuration between December 2014 and January 2015. Reduction was carried out in the same way. In addition, three stars that had been observed for the IACOBsweG catalogue with the HERMES spectrograph were found to lie in the immediate vicinity of Do~25, and are included in the analysis.  The High Efficiency and Resolution Mercator Echelle Spectrograph (HERMES) is operated at the 1.2~m Mercator Telescope (La Palma, Spain).  HERMES reaches a resolving power $R= 85\,000$, and a spectral coverage from 377 to 900~nm, though some small gaps exist beyond 800~nm \citep{raskin14}. Data were homogeneously reduced using version 4.0 of the
HermesDRS\footnote{http://www.mercator.iac.es/instruments/hermes/hermesdrs.php}
automated data reduction pipeline offered at the telescope. A complete set of bias, flat, and arc frames, obtained each night, was used to this aim. For wavelength calibration, we used a combination of a thorium-argon
lamp equipped with a red-blocking filter to cut off otherwise saturated argon lines and a neon lamp for additional lines in the near infrared. The arc images typically have a S/R in the $\sim80$\,--\,120 range. The HermesDRS pipeline provides wavelength calibrated, blaze corrected, order merged spectra. We then used our own programs developed in IDL to normalise and correct for heliocentric velocity the final spectra. The whole log of high-resolution observations is presented in Table~\ref{loghigh}.

\begin{table}
\centering
 \begin{minipage}{\columnwidth}
\caption{Log of high-resolution observations. HD~48691, HD~48807, and HD~292167 were observed with HERMES. All the other stars were observed with FIES. \label{loghigh}} 
\begin{tabular}{lcc}
\hline\hline
\noalign{\smallskip}
Star &Date of& Exposure\\ 
&observation& time (s)\\
\noalign{\smallskip}
\hline
\noalign{\smallskip}
S1 & 2011 Jan 12\,--\,13 & 5400\\
S12& 2011 Jan 12 & 5400\\
S15& 2011 Jan 15& 3600\\
S17& 2011 Jan 14& 3600\\
S33& 2011 Jan 13\,--\,14& 5400\\
HD~48691& 2013 Oct 30& 900\\
HD~48807& 2013 Oct 31& 775\\
HD~292090& 2014 Dec 9 & 900\\
HD~292392& 2014 Dec 17& 900\\
HD~292398& 2014 Dec 17& 900\\
HD~292167& 2013 Oct 30& 1300\\
HD~292163 & 2014 Dec $21 + 27$  & 3600\\
HD~292164 &  2015 Jan 5& 1800\\
\noalign{\smallskip}
\hline
\end{tabular}
\end{minipage}
\end{table}

Finally, one star, S22, was observed with the Intermediate Dispersion
Spectrograph (IDS) mounted on the 2.5-m Isaac Newton Telescope (INT; La Palma, Spain), during a run on 2011 February 24-27. We used the instrument with the 235-mm camera, the H1800V grating and the {\it Red}$+$2 CCD. S22 is the star used by \citet{lennon90} to derive abundances for Do~25. It is too faint to be observed efficiently with FIES, but we felt that a modern spectrum was needed for better comparison of results. The setup used gives a nominal dispersion of 0.34 \AA/pixel, with an unvignetted range of 750\AA. For broader spectral coverage, we took spectra at two different grating angles, centred on 4\,300\AA\ and 5\,000\AA, respectively. To compensate for poor seeing, we used a $1\farcs6$ slit, resulting in a resolution element of $\sim0.8$\AA, measured on arc lines (i.e. $R$ ranging from $\sim5\,000$ to $\sim6\,000$ across the spectral range).

%                                     Two column figure (place early!)
%______________________________________________ Gamma_1 (lg rho, lg e)
   \begin{figure}
   \centering
\resizebox{\columnwidth}{!}{\includegraphics[clip]{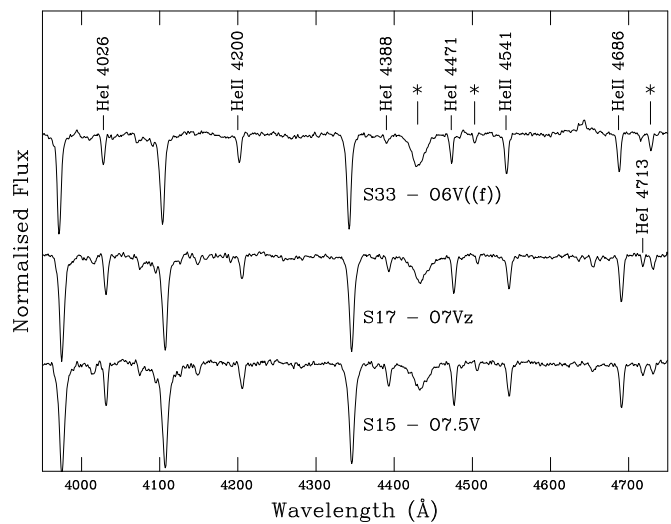}}
   \caption{Classification spectra of the earliest cluster members. Prominent spectral lines are indicated, while the strongest diffuse interstellar bands are marked with a '*'. 
              \label{ostars}}
    \end{figure}
 \section{Results}
\label{results}

In this section, we present the spectral classification of all the targets observed and abundance determinations for those observed at high resolution. For guidance, all the stars mentioned, except those discussed in Sect.~\ref{distant} (which are located far away from the cluster), are identified in Fig.~\ref{map}.

\subsection{Spectral classification of cluster members}
\label{secspec}

\begin{figure}
   \centering
\resizebox{\columnwidth}{!}{\includegraphics[clip]{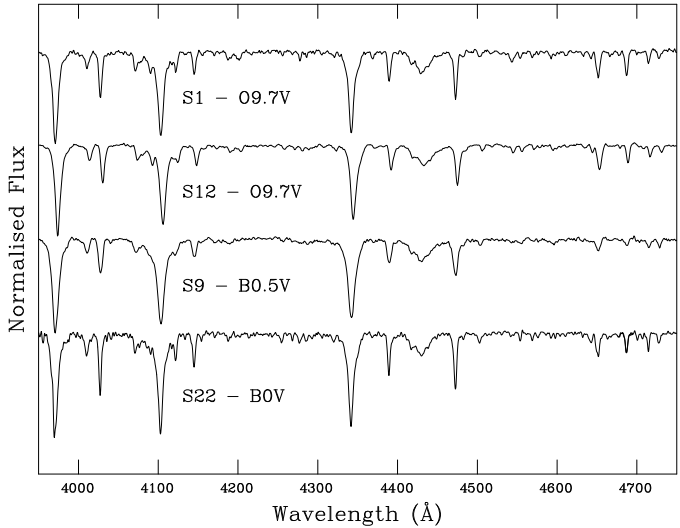}}
   \caption{Classification spectra of cluster members with spectral types around B0.
              \label{bstars}}
    \end{figure}

Classification spectra of early-type members of Do~25 are shown in Figures~\ref{ostars} and~\ref{bstars}, while their observational parameters are listed in Table~\ref{members}.

%                                     Two column figure (place early!)
%______________________________________________ Gamma_1 (lg rho, lg e)

As a first unexpected result, we can see that S15 and S17, the two O-type stars at the centre of the cluster (see Fig.~\ref{map}), are rather earlier than previously assumed \citep[in particular, than assumed by][who believed them to be O9\,V]{lennon90}. Star 17 is O7\,Vz, while S15 is O7.5\,V\footnote{The higher resolution spectrum suggests that S15 might actually be a double-lined spectroscopic binary (SB2). See Section~\ref{spctan}.}. Even more surprisingly, S33 (= TYC 148-2536-1), which lies outside the area covered by photometric studies, to the east of the cluster core, is an O6\,V star. Therefore the number of ionising photons available to excite Sh2-284 is much larger than previously assumed. All these stars are too hot to present metallic lines that can be used to derive abundances. They were, however, observed at higher resolution to determine their physical parameters.

Star 1, which lies to the north-east of Sh2-284, away from the cluster, and S12 (= BD~$+00\degr$1573s), to the north-west of the cluster, are both O9.7\,V, and were selected for high-resolution observations. The high-resolution spectra show that S12 is a SB2. Star 9, the brightest star in Cluster~2 of \citet{puga09}, is B0.5\,V, but it is a fast rotator with weak and shallow lines, and was discarded as a target for abundance determinations. Star 13 is an eclipsing binary (= GU~Mon) with two B1\,V components. It cannot be used for abundance determinations, but it is the subject of a separate study (Lorenzo et al., in prep.). Finally, S22 is B0\,V, as assumed by \citet{lennon90}, but it is too faint to be observed with FIES. Therefore it was observed at moderate resolution with the INT. Star 24 (= TYC 148-2254-1) is a Herbig Be star with emission lines. It will be discussed in Sect.~\ref{pms}.
 
%
%__________________________________________________ One column table
   \begin{table*}[ht]
      \caption[]{Observational parameters$\tablefootmark{1}$ of bright cluster members.
         \label{members}}
     $$ 
 \centering
 \begin{minipage}{\textwidth}
       \begin{tabular}{lccccccc}
            \hline
            \noalign{\smallskip}
            Star      & $U$ & $B$ & $V$& $J$ & $H$ & $K_{\mathrm{S}}$ & Spectral\\
&&&&&&&type\\
            \noalign{\smallskip}
            \hline
            \noalign{\smallskip}

S1 &11.20 &11.81 &11.38 &10.37 & 10.04 & 9.98 & O9.7\,V\\
S9 &12.35 &12.81 &12.30 &11.23 & 10.99 & 10.68& B0.5\,V\\
S12&10.50 &11.15 &10.79 &10.05 & 9.99  & 9.95 & O9.7\,V + B\\		      
S13&11.74 &12.31 &11.95 &10.89 & 10.83 & 10.76& B1\,V+B1\,V	\\	      
S15&11.87 &12.31 &11.65 &10.27 & 10.08 & 9.92 & O7.5\,V + OB?\\		      
S17&11.65 &12.07 &11.43 &10.00 & 9.79  & 9.68 & O7\,Vz\\ 		      
S22&12.05 &12.52 &12.09 &10.99 & 10.86 & 10.78& B0\,V\\		      
S24&      &12.6  &12.0  &10.32 & 9.92  & 9.48 & B0\,Ve\\	     	       
S33&      &12.2  &11.5  &9.72  & 9.44  & 9.29 & O6\,V((f))\\	    

            \noalign{\smallskip}
            \hline
         \end{tabular}
\end{minipage}
     $$ 
\begin{list}{}{}
\item[]$^{1}$ All $JHK_{\mathrm{S}}$ photometry is from 2MASS. Precision $UBV$ photometry for all stars from \citet{moffat75}. except S24 and S33, where $BV$ photometry is from UCAC4.
\end{list}
   \end{table*}
%

%
%__________________________________________________ One column table
   \begin{table*}
      \caption[]{Observational parameters$\tablefootmark{1}$ of OB stars around Dolidze~25.
         \label{around}}
      \centering
$$
 \begin{minipage}{\textwidth}
       \begin{tabular}{lccccccc}
            \hline
\hline
            \noalign{\smallskip}
            Name      & $U$ & $B$ & $V$& $J$ & $H$ & $K_{\mathrm{S}}$ & Spectral\\
&&&&&&&type\\
            \noalign{\smallskip}
            \hline
            \noalign{\smallskip}
TYC 147-2024-1   & $-$  &12.5  & 12.5 & 11.86 & 11.81 & 11.79 &B1.5\,V$+$\\    	
UCAC4 452-021858 & $-$  & $-$  & $-$  & 12.01 & 11.92 & 11.89 &B1.5\,Vn\\
HD~292167        & 9.10 &9.69  & 9.26 & 8.21  & 8.14   & 7.99  &O8.5\,Ib((f))\\ 
HD~48691         & 7.12 &7.92  & 7.83 & 7.67  & 7.73  & 7.73  &B0.5\,IV\\ 
HD~48807         & 6.83 &7.23  & 7.02 & 6.49  & 6.43  & 6.35  &B8\,Ib\\ 
HD~292090        & 9.51 &10.18 & 9.87  & 9.16 & 9.08  & 9.02  &O8\,IV\\
HD~292392        & 9.82 &10.34 & 10.01 & 9.29  & 9.21  & 9.14  &O8.5\,V\\
HD~292398        & 9.84 &10.13 & 9.92  & 9.55  & 9.54  & 9.52  &B4\,IV-V\\
HD 292163     & $-$     &11.3  & 11.0 & 10.47 &10.42 & 10.40& O8\,Vz + B\\	   
HD~292164 & $-$ &11.1 & 10.7 & 9.95 & 9.87 & 9.83& B1.5\,II-III\\	        	   
                        \noalign{\smallskip}
            \hline
         \end{tabular}
\end{minipage}
$$
\begin{list}{}{}
\item[]$^{1}$ All $JHK_{\mathrm{S}}$ photometry is from 2MASS. Precision $UBV$ photometry exists for HD~292167, HD~292090 \citep[from][]{hiltner56}, HD~48691 \citep[from][]{crampton71}, HD~292392, and HD~292398 \citep[from][]{lunel80}. For other sources, $BV$ photometry is from UCAC4. The magnitudes for TYC~147-2024-1 have huge formal errors of 0.99~mag.
\end{list}
   \end{table*}

\subsection{Stars around Dolidze~25}
\label{sec:around}

\begin{figure*}
   \centering
\resizebox{\columnwidth}{!}{\includegraphics[clip]{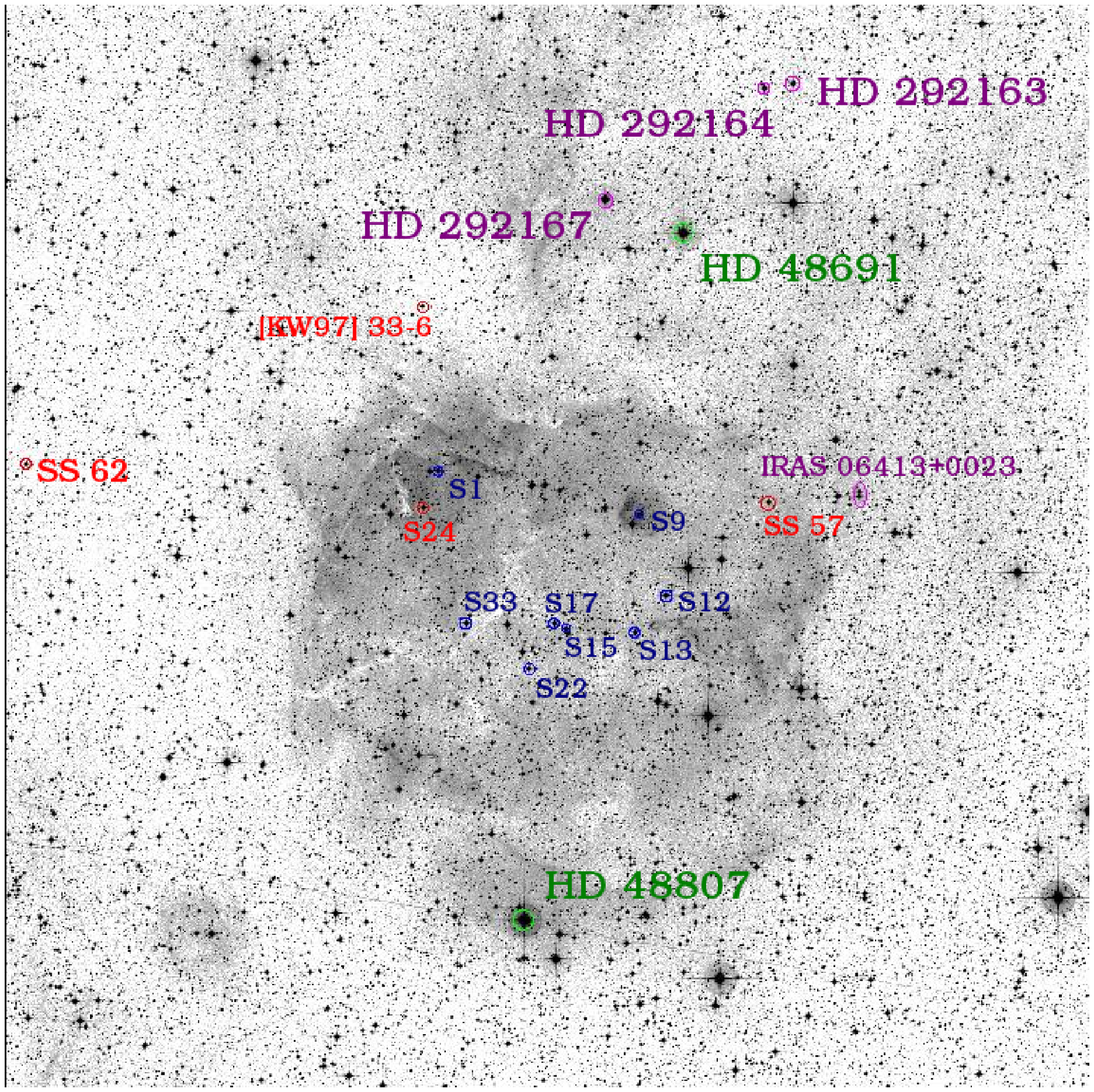}}
\resizebox{\columnwidth}{!}{\includegraphics[clip]{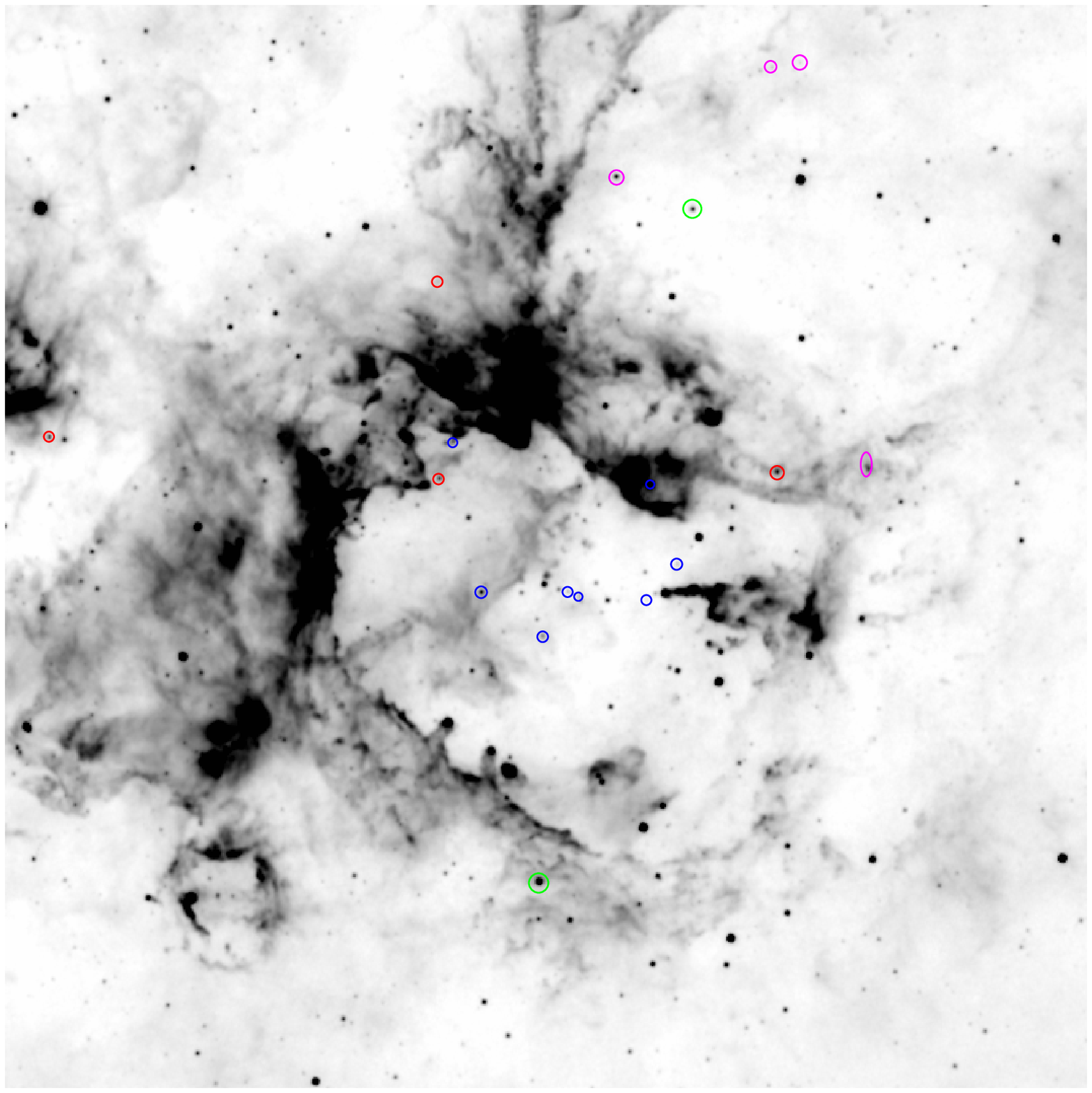}}
   \caption{Map of the area around Sh2-284. {\bf Left: } one square degree centred on the nominal position for Dolidze~25 in SIMBAD from the DSS2 red plate with the stars observed identified. Likely cluster members are indicated with their WEBDA numbers and marked in blue. They are described in Sect.~\ref{secspec}. Emission line stars are shown in red. Clearly foreground star are marked in green. Other stars at (approximately) the same distance as the cluster are marked in magenta. {\bf Right:} the same area (the centre is displaced by about $2\arcmin$ to the north and to the west) as seen by WISE in the W3 filter (12\,$\mu$m), which traces almost exclusively dust emission. The positions of all the stars in the left panel are marked with the same symbols. In both images North is up and East is left.
              \label{map}}
    \end{figure*}

Our survey extends to stars in the region that have not been traditionally considered members of Do~25. They include a number of stars towards the north-west, where images show an extension of H$\alpha$ and dust emission (see Fig.~\ref{map}). Since stars in this area seem to have properties compatible with being at the same distance as the cluster, we also included a few objects at higher distances, to probe the extent of the star-forming region. Observational parameters for all these stars are listed in Table~\ref{around}.

\subsubsection{IRAS~06413+0023}

This is a weak IRAS source on the north-west edge of the emission shell defining Sh2-284, about $18\arcmin$ from the central concentration. Two stars directly projected on top of it were classified as early-B stars by \citet{sebastian12} in a spectroscopic survey of the CoRoT fields. The mid-infrared source was too weak in the CO survey of \citet{wouter89} to measure a radial velocity, but it is clearly detected in WISE images (it can be identified as J064356.83+002039.9 or J064356.96+002051.8 in the ALLWISE catalogue, with a W4 magnitude $\sim4$~mag).

The two stars, TYC~147-2024-1 and UCAC4~452-021858, have similar spectral types, B1.5\,V, and display broad lines. TYC~147-2024-1 is likely binary, as all the lines are very asymmetric, but none of the two shows evidence of emission lines. Even though both lack reliable optical photometry, their 2MASS magnitudes suggest that they are at least as distant as Do~25. Given that IRAS~06413+0023 seems connected to the dust shell defining Sh2-284, these two stars are likely part of the young population associated with the \ion{H}{ii} region.

\subsubsection{HD~48691 and HD~48807}

These two bright stars are projected in front of Sh2-284, and have been observed as part of the IACOBsweG survey. They are included to study the run of the extinction to Do~25. HD~48691 was classified as B0.5\,IV by \citet{morgan55}, and our spectrum confirms this classification. Its spectroscopic distance modulus ($DM$) is 11~mag, corresponding to $1.6$~kpc. HD~48807 was classified as B7\,Iab by \citet{macconnell}, but our spectrum rather supports a classification B8\,Ib. This star was analysed by \citet{lefever10} with an automatic tool to determine stellar parameters of OB stars in CoRoT fields. They determine $T_{\mathrm{eff}}=12\,500$~K, $\log g = 2.0$, in good agreement with the values for B8\,Ib stars in \citet{firnstein}. Its spectroscopic distance modulus is 12.0~mag, corresponding to 2.5~kpc.. 

\subsubsection{HD~292167}

This star was classified as O9\,III: (uncertain luminosity class) by \citet{morgan55}, but appears as A3\,Ia in \citet{sebastian12}. Our spectrum indicates an O8.5\,Ib((f)) supergiant, according to the modern classification scheme of \citet{sota11}. Its spectroscopic distance is identical to that of cluster members and indeed it has been considered the main ionising source of Sh2-284 in the past \citep[e.g.][]{reynolds}. Its radial velocity is slightly different from those of cluster members. It is derived from a single measurement, so the possibility of binary motion cannot be ruled out.

\subsubsection{HD~292090, HD~292392, and HD~292398}
\label{distant}

These three stars are located at moderate distances from Sh2-284, outside the area shown in Fig.~\ref{map}, but their photometric colours suggested they could be OB stars at the same distance. HD~292398, however, is clearly a foreground star. Even though its inclusion in the Luminous Star Catalog (LS VI $+00^{\degr}29$) together with its colours suggested a late-B supergiant, it is not very luminous. We derive a spectral type B4\,IV--V. Its spectroscopic distance modulus is $\approx10.5.$, i.e. $d \approx 1.3$~kpc.

HD~292090 was classified as O8 by \citet{morgan55}, and B3\,Ib by  \citet{sebastian12}. Even though our spectrum is quite noisy, it is clearly a late-O star. In principle, we would classify it as O8\,IV, though the poor spectrum allows for a slightly earlier or later type and a III or V luminosity class. We choose class IV because the wind parameter suggests that it is not a main sequence star and, as discussed in Section~\ref{disc}, indirect evidence suggests that it is at a distance similar to that of Do~25. With this classification, its distance modulus is 12.9 ($d=3.8$~kpc). HD~292392 was classified as O8n in the McDonald system by \citet{popper44}. Two radial velocity measurements disagree by almost $20\:\mathrm{km}\,\mathrm{s}^{-1}$, but the average value does not differ much from other stars in the region \citep{popper44}. We obtain a spectral type O8.5\,V. with a corresponding distance modulus 12.3 ($d=2.9$~kpc).

\subsubsection{HD~292163 and HD~292164}

Though SIMBAD assigns to LS\,VI~$+00^{\degr}13$ and LS\,VI~$+00^{\degr}14$ positions that do not coincide exactly with any star, the identifications HD 292163 =  LS\,VI~$+00^{\degr}13$ and HD 292164 = TYC 147-1684-1 = BD~$+00^{\degr}$1569 = LS\,VI~$+00^{\degr}14$ seem immediate. In spite of the spectral type G0 assigned to both in the database, their early types are confirmed by their spectra.

HD~292163 was observed twice, and the two spectra show a large change in radial velocity. The spectrum with the best S/R shows an O8\,V star and a second component, corresponding to a spectral type not far from B0\,V.  HD~292164 is a B-type star with very narrow lines. We derive a spectral type B1.5\,II-III. The two stars are very close together in the sky (separated by $\sim1\farcm5$), have similar, very low colour excesses $E(J-K_{\mathrm{S}})\simeq0.27$, and almost identical interstellar lines. However, they cannot belong to the same population, as the less massive star would be the most evolved. Both stars have spectroscopic distances compatible with Do~25, but HD~292163 could be more distant if the second component contributes significantly to the total brightness.

\subsection{High-mass PMS stars}
\label{pms}

\begin{figure}
   \centering
\resizebox{\columnwidth}{!}{\includegraphics[clip]{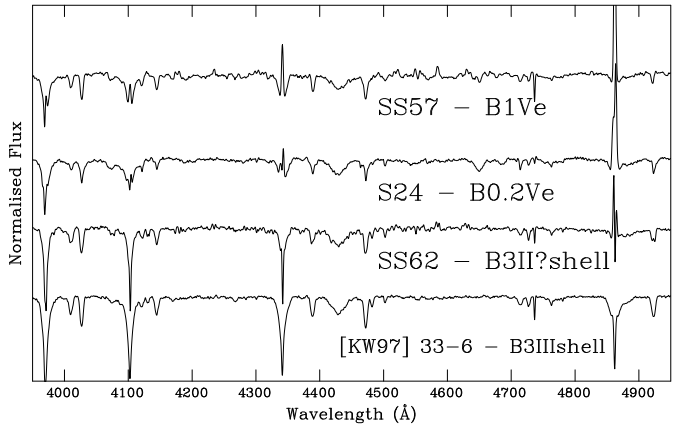}}
   \caption{Classification spectra of emission-line stars in the area of Do~25. The narrow feature 4720\AA\ is due to a defect of the CCD.
              \label{herbigs}}
    \end{figure}

A number of emission-line stars in the area have magnitudes and colours compatible with membership in Do~25 or the surrounding star-forming region. Since their spectra do not allow the derivation of stellar parameters, we discuss their main characteristics. Spectra are shown in Fig.~\ref{herbigs}, while infrared photometry is displayed in Table~\ref{bes}.

%                                     Two column figure (place early!)
%______________________________________________ Gamma_1 (lg rho, lg e)
 
%
%__________________________________________________ One column table
   \begin{table*}
      \caption[]{Infrared photometry of emission-line stars$^{1}$, and S9.
         \label{bes}}
     $$ 
 \centering
 \begin{minipage}{\textwidth}
       \begin{tabular}{lcccccccc}
            \hline
\hline
            \noalign{\smallskip}
            Name & $J$ &$H$ & $K_{\mathrm{S}}$ & W1 & W2 & W3 & W4 &Spectral\\
&&&&&&&&type\\
            \noalign{\smallskip}
            \hline
            \noalign{\smallskip}
            S9$^{2}$    & 11.23&10.99 &10.68& 9.97& 9.53 & 8.02& 4.04&B0.5\,V\\
            S24   & 10.32& 9.92 & 9.48& 9.52 &9.26 & 8.55& 5.94&B0\,Ve\\ 
            SS 57 & 10.47&10.17 & 9.89& 9.43 &9.08 & 7.96& 5.91&B1\,Ve\\
            SS 62 & 10.47&10.12 & 9.81& 9.56 &9.26 & 8.43& 6.43&B3\,II:shell\\
            $\left[ \mathrm{KW97}\right]$ 33-6& 11.53&11.31 &11.18 & 10.97&10.86 & 11.41$^{3}$&$-$& B3\,IIIshell\\
                        \noalign{\smallskip}
            \hline
         \end{tabular}
\end{minipage}
     $$ 
\begin{list}{}{}
\item[]$^{1}$ All $JHK_{\mathrm{S}}$ photometry is from 2MASS. Mid-infrared photometry is from the ALLWISE catalogue.
\item[]$^{2}$ S9 is not an emission-line star, but it is immersed in the bright nebulosity of IRAS 06422+0023. Its mid-infrared colours are thus dominated by nebular emission, and serve as a comparison. 
\item[]$^{3}$ This value is not reliable, as its associated error is $\pm0.52$
\end{list}
   \end{table*}

\subsubsection{S24}

Star 24 = TYC~148-2254-1 is an emission-line star located on the emission shell defining Sh2-284, about $9\arcmin$ north-east of the central condensation. It has a very early spectral type B0\,Ve.  All emission-lines are strong and double-peaked. In 2009, they present a much stronger blue peak, but by 2014 the symmetry had reversed: emission lines have a broad blue component and a very narrow sharp red component (spectrum displayed in Fig.~\ref{herbigs}). H$\alpha$ displays an equivalent width (EW) of $-25\pm1$\AA, and the \ion{He}{i}~6678\AA\ and~7065\AA\ lines are also in emission. This star was studied by \citet{semaan13}, as part of an investigation of Be stars in CoRoT fields, but they could not derive stellar parameters because of the strong veiling.

According to \citet{semaan13}, the location of S24 in the $(J-H)_{0}/(H-K_{\mathrm{S}})_{0}$ diagram is typical of an extreme classical Be star. It has the strongest near-infrared excess of all their sample, lying close to the Herbig Be star region. Given the mid-IR excess (W1$-$W3 $\approx 1$~mag; see Table~\ref{bes}) and its location in an active star-forming shell, we consider it is very likely a Herbig Be star. Classical Be stars, on the other hand, are extremely rare (or even completely absent) in very young open clusters \citep[e.g.][]{mathew08}.

\subsubsection{SS 57}

SS~57 = TYC 147-506-1 is an emission-line star located on the emission shell defining Sh2-284, about $13\arcmin$ north-west of the central condensation. It has a spectral type B1\,Ve, and a strong near-IR excess (see Fig.~\ref{CMD}). Indeed, it is clearly located within a local maximum of the dust emission, as seen in WISE W4 images, and also displays a very strong mid-IR excess (W1$-$W3 $\approx 1.5$~mag; see Table~\ref{bes}). All emission-lines are strong and single-peaked. H$\alpha$ displays an EW of $-53\pm2$\AA, but the \ion{He}{i} lines are all strongly in absorption. This star was studied by \citet{semaan13}, as part of a study of Be stars in COROT fields. They confirm the B1\,V spectral type and derive parameters typical of a $\sim11\:M_{\sun}$ star on the ZAMS.

Again in this case \citet{semaan13} indicate that the location of SS~57 in the $(J-H)_{0}/(H-K_{\mathrm{S}})_{0}$ diagram is typical of an extreme classical Be star rather than a Herbig Be star. However, the strong mid-IR excess and its location in an active star-forming shell make it a very strong candidate for a Herbig Be star in Sh2-284.

Since both S24 and SS~57 lie within H$\alpha$ nebulosity, it is worthwhile checking whether nebular emission contributes significantly to their emission features. However, inspection of the raw spectra does not reveal any increase in the intensity of the sky lines in the immediate vicinity of any of the stars. Moreover, there does not seem to be H$\alpha$ knots associated with them.

\begin{figure*}
   \centering
\resizebox{\textwidth}{!}{\includegraphics[clip,angle=+90]{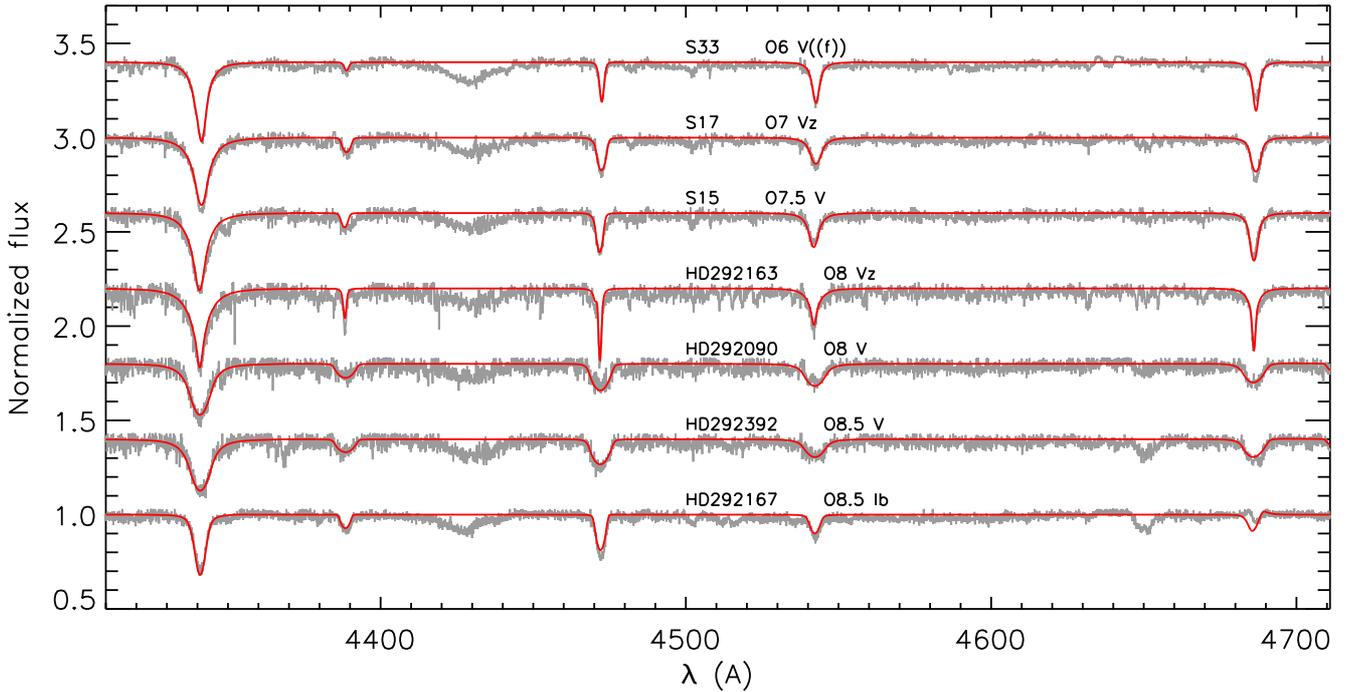}}
   \caption{Observed spectra (think grey line) and best-fit model (thin red line) for a sample of O-type stars. We note that the metallic features (\ion{Si}{iv} and \ion{C}{iii}) have not been fitted. The spectrum of HD~292090 is of very low quality, and this is reflected in the parameters obtained in Table~\ref{parameters}.\label{fitO}}
    \end{figure*}

\subsubsection{SS 62}

SS~62 is located about $17\arcmin$ to the north-east of the shell defining Sh2-284, just on the edge of a smaller dust-emission shell that is not readily visible in H$\alpha$. This shell seems to be associated with the strong IRAS point source IRAS~06446+0029 = MSX6C~G212.0641$-$00.7395, which has exactly the same radial velocity as the IRAS sources directly associated with Sh2-284 \citep[see Table~\ref{iras}]{wouter89,bronfman96}. Our spectrum is very peculiar. The Balmer lines display shell-like features, with strong blue emission peaks in both H$\alpha$ and H$\beta$. Most \ion{He}{i} lines are very asymmetric, perhaps also due to emission components. The apparently photospheric features and the wings of the higher Balmer lines, which do not seem affected by emission, correspond to a luminous B3 star. Judging from the width of the Balmer lines, under the assumption that the emission components only affect the line cores (as seems the case for H$\gamma$ and H$\delta$), the luminosity class is around II. This is a very high luminosity for a classical Be star \citep{negueruela04}. Considering the association with an IRAS source and strong mid-infrared (W1$-$W3 $\approx 1.1$~mag; see Table~\ref{bes}), this object is very likely a Herbig Be star, even if the strength of the H$\alpha$ emission is moderate ($EW=-31\pm2$\AA).

\subsubsection{$\left[ \mathrm{KW97}\right]$ 33-6}

The emission-line star $\left[ \mathrm{KW97}\right]$ 33-6 lies to the north of Do~25, a few arcmin beyond the edge of the \ion{H}{ii} shell. The WISE images show some diffuse dust emission in the area, but no point source associated with this object. The spectrum shows a B3\,III star with weak shell characteristics. H$\alpha$ has a typical emission peak with a central self-absorption going below the continuum, and $EW=-3.5\pm0.5$. Given the weak emission characteristics and the lack of mid-infrared excess (the object is not even detected in W4), it is likely to be a classical Be star, not associated with the cluster. However, we note that, unless its photometry is severely affected by the circumstellar disk, it suffers heavy extinction, and its distance modulus suggests that it is located beyond the cluster, at $d=5$\,--\,6~kpc. 

\subsection{Spectroscopic parameters and abundances}
\label{spctan}

Stellar and wind parameters for the sample of O-type stars observed at high resolution were determined using the {\sc iacob-gbat} package \citep{ssimon11}, based on a $\chi^2$-fitting algorithm applied to a large pre-computed grid of {\sc fastwind} \citep{santol97,puls05} synthetic spectra, and standard techniques for the hydrogen/helium analysis of O-type stars \citep[see e.g.][]{herrero92,repolust04}. We refer the reader to \citet{sabin14} for a detailed description of the strategy followed. In this case we used the grid of {\sc fastwind} models computed for solar metallicity. The line-broadening parameters (projected rotational velocity, $v\sin i$, and macroturbulent broadening parameter, $v_{\rm mac}$) used as input for the {\sc iacob-gbat} analysis were previously computed by applying the {\sc iacob-broad} tool \citep{ssimon14} to the \ion{O}{iii}\,5591\AA\ line. Table~\ref{parameters} summarises the resulting parameters, also including estimates for the associated uncertainties. In a few cases, {\sc iacob-gbat} only provided upper or lower limits for some parameters. While the upper limits in the wind-strength parameter\footnote{The wind-strength parameter groups together the mass-loss rate ($\dot{M}$), the stellar radius ($R$) and the wind terminal velocity ($v_{\infty}$) into the expresion $Q$\,=\,$\frac{\dot{M}}{(Rv_{\infty})^{1.5}}$} found for S15, S17, and HD\,292163 are simply an indication of the lack of sensitivity of the H$\alpha$ and \ion{He}{ii}\,4686\AA\ lines (main diagnostic lines in the optical spectral range of the stellar wind properties) below $\log Q\simeq-13.5$, the lower limits in $\log g$ found for S15 and HD\,292163 are very likely showing that we are dealing with composite spectra\footnote{This is certainly the case for HD\,292163, which is an SB2, and likely for S15 too, as its lines show evidence of a secondary component and its radial velocity is somewhat different from the cluster average (most clearly seen in Fig.~\ref{redhelium}).}. The best-fit models (containing only H and He lines) are displayed in Fig.~\ref{fitO}, compared to the observed spectra. In the case of HD~292167, the fit is far from perfect. As can be seen in Fig.~\ref{fitO}, there is a moderately strong emission component in the blue wing of \ion{He}{ii}~4686\AA, one of the main wind diagnostics, instead of the expected P-Cygni or pure emission profile. The model cannot fit simultaneously this line and H$\alpha$, and the wind parameters derived are a compromise between the two lines. This sort of discrepancy is not unusual in O-type supergiants \citep[see e.g. the case of HD~192639 and HD~193514 in][]{repolust04}. This behaviour may be due to wind clumping or, perhaps, to some complex dynamical behaviour at the base of the wind, but its characterisation clearly lies beyond the scope of this paper.

\begin{table*}
 \centering
 \begin{minipage}{\textwidth}
  \caption{Stellar parameters derived from the \textsc{fastwind} analysis for the stars that are too hot to derive abundances.\label{parameters}}
  \begin{tabular}{lccccccccc}
  \hline
\hline
            \noalign{\smallskip}
Name   & Spectral &$v \sin i$&$v_{\mathrm{mac}}$ & $v_{\mathrm{rad}}$ & $T_{\mathrm{eff}}$ & $\log g$ & $Y_{\mathrm{He}}$ & $\xi$ & $\log Q$   \\
& type &($\mathrm{km}\,\mathrm{s}^{-1}$) &($\mathrm{km}\,\mathrm{s}^{-1}$)& ($\mathrm{km}\,\mathrm{s}^{-1}$)& (kK)& (dex) & & ($\mathrm{km}\,\mathrm{s}^{-1}$) & \\
 \noalign{\smallskip}
 \hline
 \noalign{\smallskip}
S15 & O7.5\,V&105 &   40  & 36&   $38.8\pm0.8$ &  $>4.16$  &  $<0.10$ & 5 & $<-13.5$\\
S17 & O7\,Vz&140 &    0  & 45 &  $38.1\pm1.0$ &  $4.1\pm0.2$ & $0.12\pm0.02$ & 5 & $<-13.5$\\
S33 & O6\,V((f))& 53 & 76 & 48 & $40.0\pm1.0$ &  $3.9\pm0.1$ & $0.11\pm0.02$ & 10 & $-12.8\pm0.2$\\
HD~292163 & O8\,Vz+& 35 & 0 &4 (var)& $39.2\pm0.8$ & $>4.2$ & $0.20\pm0.05$ & 5 & $<-13.5$\\
HD~292090 & $\sim$O8\,V & 280 & 0 & 4: & $36.6\pm2.8$ & $3.9\pm0.4$ & $0.1\pm0.1$ & 10& $-13.3$\\
HD~292392 & O8.5\,V & 280 & 0 & 29 & $35.8\pm2.3$ & $3.9\pm0.4$ & $0.21\pm0.09$ & 10 & $-12.8\pm0.2$\\ 
HD~292167 & O8.5\,Ib((f)) & 150 & 0  & 31 & $32.6\pm1.7$ & $3.3\pm0.2$ & $0.12\pm0.04$ & 10 & $-12.3\pm0.1$\\
 \noalign{\smallskip}
\hline
\end{tabular}
\begin{list}{}{}
\item[] The typical uncertainty for $\xi$ is $2:\mathrm{km}\,\mathrm{s}^{-1}$.
\end{list}
\end{minipage}
\end{table*}

\begin{figure*}
   \centering
\resizebox{\textwidth}{!}{\includegraphics[clip,angle=+90]{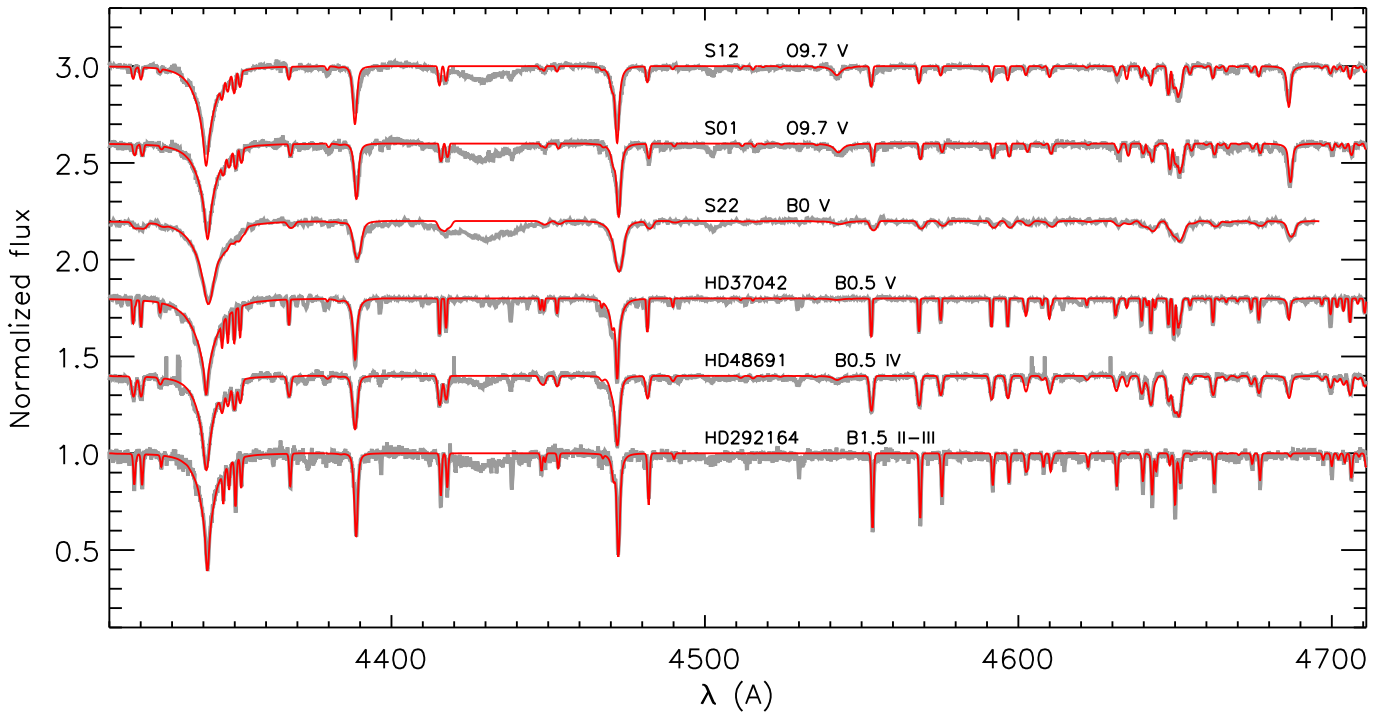}}
   \caption{Observed spectra (think grey line) and best-fit model (thin red line) for a sample of stars with strong metallic lines. The bright star HD~37042 has been included as reference. We note that the spectrum of S22 has lower resolution and a smaller wavelength coverage.
              \label{fitB}}
    \end{figure*}

\begin{figure*}
   \centering
\resizebox{\textwidth}{!}{\includegraphics[clip]{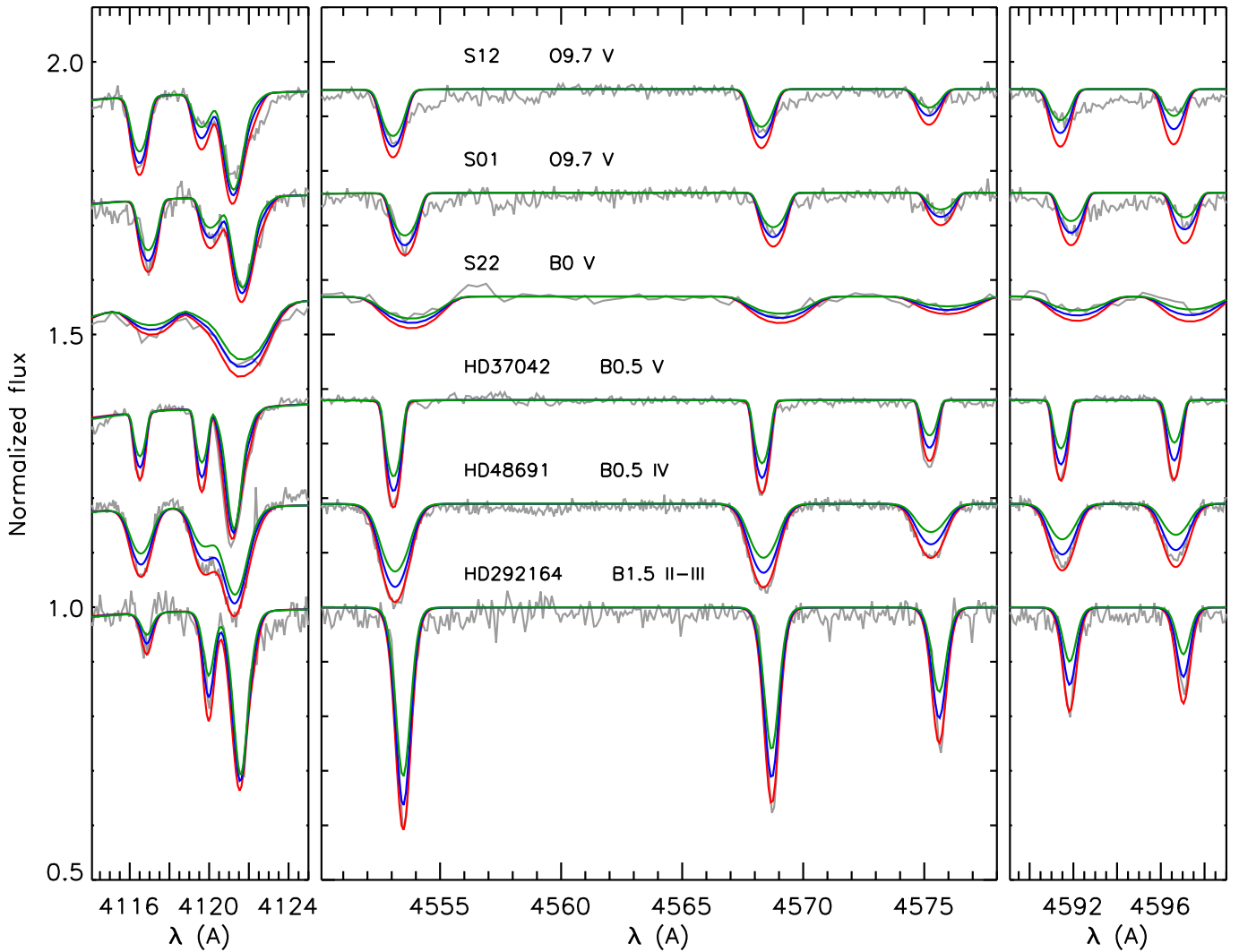}}
   \caption{Abundance analysis for the stars listed in Table~\ref{parameters_abun}, plus S22, the star used by previous authors to derive abundances in Do~25. The left panel shows the region of the \ion{Si}{iv}~4116\AA\ line; the middle panel shows the \ion{Si}{iii} triplet, while the right panel shows two representative \ion{O}{ii} lines. For each star, we display three models, calculated using the same stellar parameters, but with different abundances. The red model uses solar abundances (8.75~dex for O and 7.50~dex for Si); the blue model uses abundances 0.3~dex lower, while the green model uses abundances 0.6~dex below solar.
              \label{fit_abun}}
    \end{figure*}

Stellar parameters, along with the silicon and oxygen abundances, for the stars in our sample showing prominent metallic lines (S1, S12, and HD\,292164) were obtained by applying the strategy described in \citet{ssimon10} to the corresponding high-resolution spectra. In brief, the stellar parameters were derived by comparing the observed H (Balmer) line profiles and the ratio of \ion{Si}{iii-iv} line equivalent widths (EWs) with the output from a grid of {\sc fastwind} models computed to this aim. Then, the same grid of models was used to derive the stellar abundances by means of the curve-of-growth method. As a final sanity check, we overplotted the synthetic spectra corresponding to the resulting parameters and abundances to the observed ones (see Fig.~\ref{fitB}). This last step was especilly important given the lower quality (in terms of signal-to-noise ratio) of some of the spectra used here with respect to those analysed in \citet{ssimon10}. In all cases, we assumed a normal helium abundance ($Y_{\rm He}=0.10$) and fixed the wind-strength parameter to $\log Q = -13.5$. For comparative purposes, we also analysed in the same way the spectra of HD\,37042 (B0.5\,V), a star in the  solar neighbourhood, and HD\,48691 (B0.5\,IV), a foreground star along the same line of sight. A summary of results is presented in Table~\ref{parameters_abun} and a illustrative plot overplotting {\sc fastwind} synthetic spectra with fixed parameters (per star) and varying abundances to the observed ones is presented in Fig.~\ref{fit_abun}. 

The star HD~48807 is not a cluster members, and its temperature is too low for this grid of models. We calculated its parameters using the automatised $\chi^{2}$ fitting of synthetic \textsc{Fastwind} spectra to the global
spectrum between 3900 and 5100\AA, as described in \citet{castro12}. We find $T_{\textrm{eff}}=12\,000\:$K and $\log g =2.0$, in excellent agreement with the parameters found by \citet{lefever10}.

Finally, while it was not possible to perform a similar abundance analysis for S22 (B0\,V) because of the poorer quality of the spectrum (in terms of resolution) and because this star has a higher projected rotational velocity, we include it in Figures~\ref{fitB} and~\ref{fit_abun} for comparison, since the abundances derived by \citet{lennon90} and \citet{rolleston00} are based on the analysis of this star alone. We carried out a spectral synthesis analysis by eye, using the ratio of \ion{Si}{iii} to \ion{Si}{iv} lines, the \ion{He}{i} lines and the wings of H$\gamma$ to estimate approximate parameters for S22, finding values $T_{\textrm{eff}}\approx30\,000$~K, $\log g \approx 4.1$ and $v \sin i\approx 110\:\textrm{km}\,\textrm{s}^{-1}$. The models displayed in Figures~\ref{fitB} and~\ref{fit_abun} correspond to these parameters. In Fig.~\ref{fit_abun}, the same three abundances are used as for the other stars.

\begin{table*}
 \centering
 \begin{minipage}{\textwidth}
  \caption{Stellar parameters and abundances derived from the \textsc{fastwind} analysis for stars with prominent metallic lines. Parameters for the bright star HD~37042 are included for comparison with the solar neighbourhood. \label{parameters_abun}}
  \begin{tabular}{lccccccccc}
  \hline
\hline
            \noalign{\smallskip}
Name   & Spectral &$v \sin i$&$v_{\mathrm{mac}}$ & $v_{\mathrm{LSR}}$ &$T_{\mathrm{eff}}$ & $\log g$ & $\xi$ & $\epsilon$(Si) & $\epsilon$(O)   \\
& type &($\mathrm{km}\,\mathrm{s}^{-1}$)& ($\mathrm{km}\,\mathrm{s}^{-1}$)& ($\mathrm{km}\,\mathrm{s}^{-1}$)& (K)& (dex)& ($\mathrm{km}\,\mathrm{s}^{-1}$) & (dex)& (dex)\\
&&&&$\pm5\:\mathrm{km}\,\mathrm{s}^{-1}$&$\pm1\,000$& $\pm0.1$ &$\pm2.0$& $\pm0.15$& $\pm0.15$\\
 \noalign{\smallskip}
 \hline
 \noalign{\smallskip}
S1 &  O9.7\,V &50 &  10  &  49&  $31\,000$ &   $4.1$ & 1.0 &7.25 & 8.25\\ 
S12 & O9.7\,V& 45 &  20  &  19&  $30\,000$ &  $4.1$  & 3.0 &7.30 &8.25\\
HD~37042 & B0.5\,V &  31 &   0& 17& $29\,000$& $4.2$ & 1.5 &7.55 &8.75\\
HD~48691 & B0.5\,IV&  54 &  47 &  17&$27\,500$& $3.6$& 7.5 &7.50 &8.76\\
HD~292164& B1.5\,II-III & 17&33 &45&$22\,000$&$3.3$ & 9.0  &7.40 &8.66\\    
 \noalign{\smallskip}
\hline
\end{tabular}
%\begin{list}{}{}
%\item[] Typical uncertainties are
%\end{list}
\end{minipage}
\end{table*}

For HD\,37042 and HD\,48691, we obtained silicon and oxygen abundances compatible with those obtained for other stars in the solar vicinity \citep{przyb08,ssimon10,nieva11,nieva12}, out to distances up to 1~kpc \citep[similar to the distance to HD~48691]{garrojas14}. The spectroscopic analysis of S1 and S12 results in abundances $\sim$0.2\,--\,0.3 dex (Si) and $\sim0.5$ dex (O) below the reference stars (Table~\ref{parameters_abun}). The spectrum of S22 is compatible with similarly low abundances (Fig.~\ref{fit_abun}). Solar abundances are clearly too high for this object, but it is difficult to choose between models with $-0.3$~dex and $-0.6$~dex because of the lower quality of the spectrum and higher rotational velocity. We can, in any case, conclude that its abundance is compatible with those found for S1 and S12. Though these values are low for Galactic stars, in all these cases they seem to indicate that the metallicity in the region (at least, using Si and O as proxies of the metallicity) is not as low as previously assumed \citep[one-sixth solar][]{lennon90,rolleston00}. Interestingly, HD\,292164 that seems to be at the same distance as Do~25, has rather higher abundances, in between the reference stars and the other three targets.

As a final remark, we indicate that S12 displays clear signatures of a companion (see the Si and O lines in Fig.~\ref{fit_abun}). The presence of the companion is unlikely to have an effect on the temperature derived, as \ion{Si}{iv} is not affected, and the \ion{Si}{iii} lines can be fitted well. It may have an effect on the effective gravity, but this has been taken into account, and the fit to the blue wings has been always preferred. The abundances derived (under the assumption that the star is single) should be slightly revised upwards to take into account the dilution of the spectral features in the combined continuum.

\section{Discussion}
\label{disc}

We have presented the first spectroscopic survey of Dolidze~25 and its vicinity, and carried out an abundance analysis for two O9.7\,V members, and a B1.5\,III giant in its vicinity. As discussed in previous works, the issues of distance, metallicity and cluster parameters are intimately linked for Do~25. In the following sections, we will address them in turn, even though our conclusions are drawn by considering all of them at the same time.

\subsection{Distance to the cluster}

A first approximation to the distance to the cluster can be obtained by calculating the spectroscopic distances of all the members with spectral types. Spectroscopic moduli calculated with the optical and the infrared photometry agree very well, certifying the assumption of standard reddening. The raw average is $DM =13.2$. If we remove the binaries S12 and S15 (certain and likely, respectively), which should be intrinsically brighter than corresponds to single stars and hence appear less distant than obtained with this simple method, the average is $DM=13.3$ with a standard deviation of only $0.2$~mag, which is much lower than the expected internal scatter of the calibrations. This distance modulus corresponds to 4.6~kpc. We note two effects that can lead to a much larger uncertainty. Firstly, the calibration of \citet{turner80}, as all empirical absolute magnitude calibrations, is based on stars from the solar neighbourhood that have higher metallicities than members of Do~25. This should lead to a reduction in the distance, as discussed in the Introduction. On the other hand, there is a rather high probability that many of the stars observed are really unresolved binaries and thus intrinsically brighter than we are assuming.

\begin{figure}
   \centering
\resizebox{\columnwidth}{!}{\includegraphics[clip]{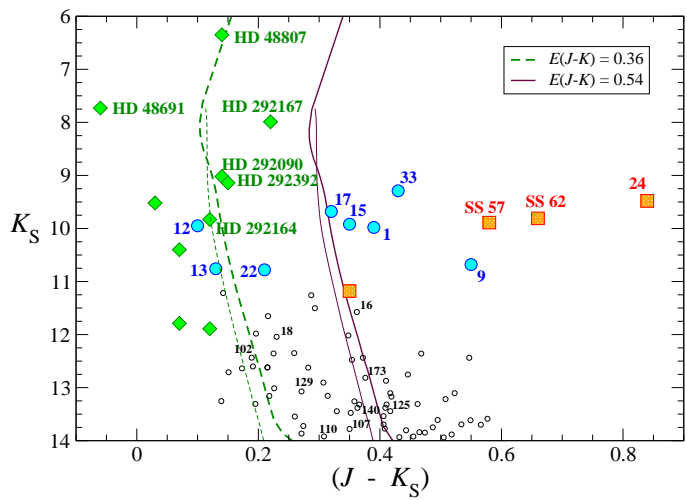}}
   \caption{Composite observational CMD for Dolidze~25 and stars in its neighbourhood. Small open circles represent likely cluster members selected from 2MASS. Accompanying numbers indicate that the star is also a photometric member in \citet{turbide} or \citet{delgado10}. Large filled (blue) circles are members with spectroscopy. (Green) diamonds represent stars in the cluster neighbourhood. The emission-line stars discussed in Section~\ref{pms} are represented by squares. The lines are 2~Ma Geneva \citep{ekstrom12,georgy13low} isochrones for fast rotation at $Z=0.014$ (thick lines) and $Z=0.002$, displaced to $DM = 13.2$, with different amounts of extinction (see text). For clarity star numbers from WEBDA are shown without any prefix.
              \label{CMD}}
    \end{figure}

The effect of lower metalicity in the distances estimated can be seen in Fig.~\ref{CMD}, where isochrones from \citet{ekstrom12} for solar metallicity and \citet{georgy13low} for $Z=0.002$ (i.e. $-0.8$~dex) are shown. The low-metallicity isochrones are very slightly bluer\footnote{Less than 0.02~mag in $(J-K_{\textrm{S}})_{0}$ and $(B-V)_{0}$.}, but a star of a given mass is fainter by $\approx0.4$~mag. As these effects are intrincately linked, we will try to constrain the distance more strongly by resorting to indirect methods.

\begin{figure}
   \centering
\resizebox{\columnwidth}{!}{\includegraphics[clip]{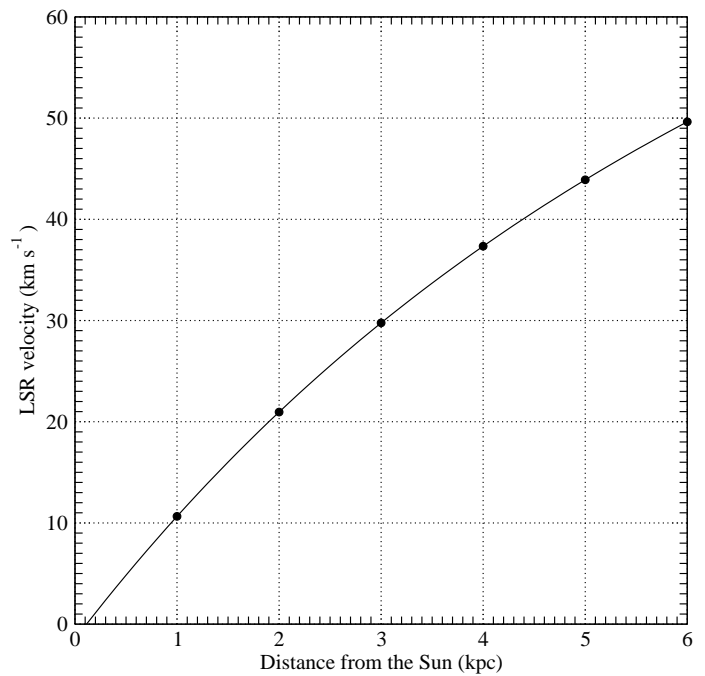}}
   \caption{Radial velocity with respect to the local standard of rest (LSR) due to Galactic rotation as a function of distance in the direction to Do~25, calculated using the parameters of \citet{reid14}.
              \label{RotCurve}}
    \end{figure}

\citet{reynolds} found strong H$\alpha$ emission at  $v_{\mathrm{LSR}}=+44.0\:{\rm km}\,{\rm s}^{-1}$ (velocity with respect to the local standard of rest) associated with Sh2-284. In an H$\alpha$ survey using a Fabry-Perot spectrometer with an aperture of $2\arcmin$ and a spectral resolution of $15\:{\rm km}\,{\rm s}^{-1}$  \citet{fich90} found emisison towards Sh2-284 composed of a foreground contribution at $v_{\mathrm{LSR}}=-20.1\pm0.5\:\mathrm{km}\,{\rm s}^{-1}$, and a stronger broad component centred on $+43.0\pm0.5\:{\rm km}\,{\rm s}^{-1}$ with a line width of $49\:{\rm km}\,{\rm s}^{-1}$. This is in agreement with measurements of CO that give a central velocity  $v_{{\rm LSR}}=+45.0\pm0.7\:{\rm km}\,{\rm s}^{-1}$ and a line width of  $2.7\:{\rm km}\,{\rm s}^{-1}$ \citep{blitz82}. In Fig.~\ref{RotCurve}, we show the dependence of radial velocity with distance along this line of sight, according to the Galactic rotation curve derived by \citet{reid14} from parallax distances to sources displaying masers. According to this curve, a velocity of $+43.0\:{\rm km}\,{\rm s}^{-1}$ corresponds to a distance of $\approx4.9$~kpc, while  $+45.0\:{\rm km}\,{\rm s}^{-1}$ corresponds to $\approx5.2$~kpc.

\begin{figure*}
   \centering
\resizebox{\textwidth}{!}{\includegraphics[clip]{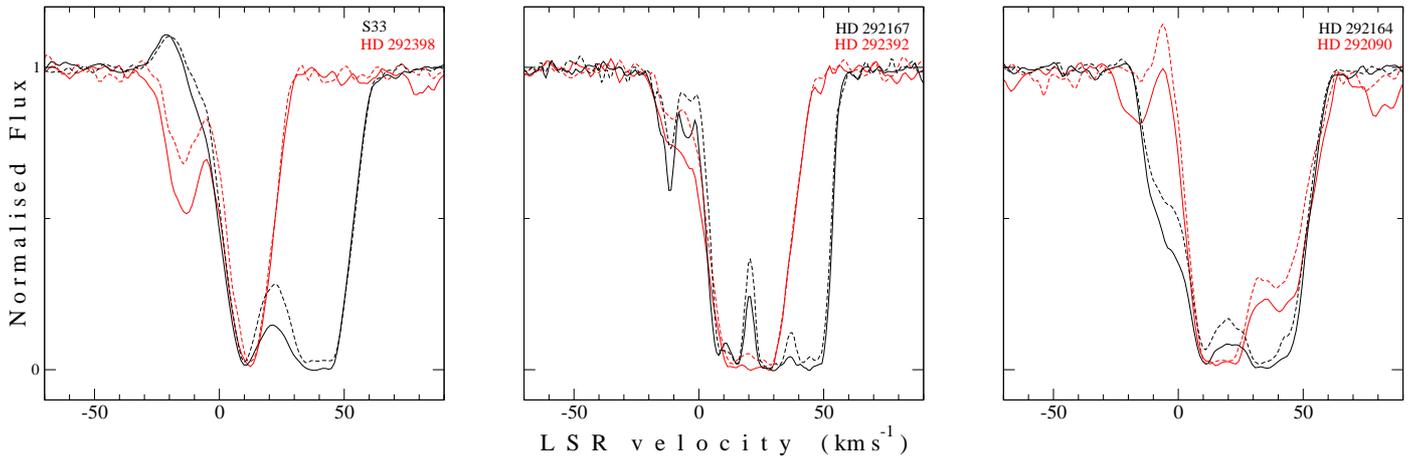}}
   \caption{Montage of the two \ion{Na}{i} D interstellar lines in velocity space for a sample of stars in and around Do~25. In each panel, a star at the distance of the cluster is shown in black, while a foreground star is shown in red. The solid line is the D1 lines, while the dashed line is the D2 line. See text for details.
              \label{NaD}}
    \end{figure*}

\begin{figure}[ht!]
   \centering
\resizebox{\columnwidth}{!}{\includegraphics[clip]{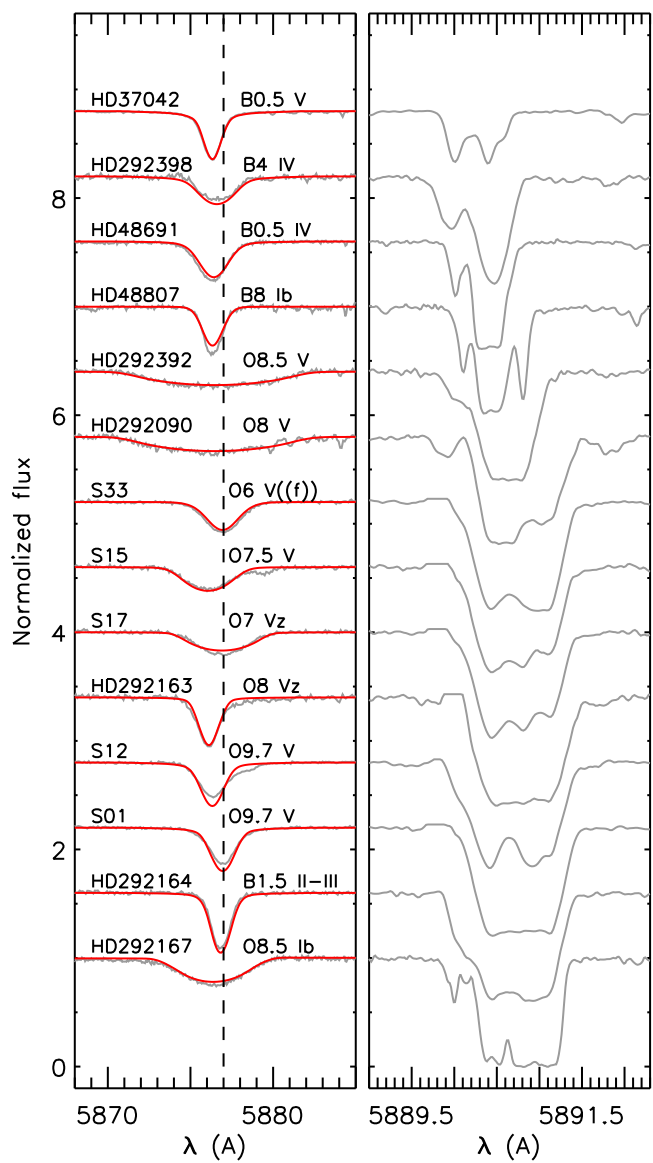}}
   \caption{The {\bf left panel} shows the observed (thin gray line) and modelled (red line) profile of the \ion{He}{i}~5875\AA\ line for all the stars analysed. The vertical line shows the rest position of the line displaced to the $v_{\mathrm{hel}}$ velocity of the gas in Sh2-284, which coincides within observational errors with the measured velocity of S1, S17, and S33, and HD\,292764. The {\bf right panel} shows the shape of the interstellar \ion{Na}{i} D1 line for each star. The different components of the line are probing the velocity of the intervening medium. 
              \label{redhelium}}
    \end{figure}

Another way of estimating the distance to the cluster is by using the different kinetic components of the interstellar lines, which are sampling the distribution of material along the line of sight.
In Fig.~\ref{NaD} we show the interstellar \ion{Na}{i} D lines for six stars observed with FIES in and around the cluster. Other interstellar lines in their spectra have a similar structure. The left panel shows S33, the earliest (and second most reddened) cluster member, and the foreground star HD~292398. The middle panel shows HD~292167, an O8.5\,Ib supergiant to the north of the cluster, compared to HD~292392, an O8.5\,V star with a shorter spectroscopic distance. Finally, the right panel shows HD~292164, a B1.5\,II-III star about $30\arcmin$ north-west of the cluster, compared to HD~292090, an O-type star $\sim90\arcmin$ south-west of the cluster. The most remarkable aspect of the line analysis is the very strong similarity in the structure of the interstellar lines for most of the stars, except that in stars with low S/N (S33 and HD~292090) the telluric lines have not been properly subtracted, and there is an artefact resembling an emission component. For reference, one of \ion{Na}{i} D lines is shown for all the stars observed at high resolution in Fig.~\ref{redhelium}.

All the stars with good S/N show a weak component at negative velocities, ranging from $-5$ to $-15\:{\rm km}\,{\rm s}^{-1}$, depending on the position on the sky. These components almost certainly arise from material very close to the Sun, given their variability from source to source and that the Galactic rotation curve does not include any negative velocities in this direction. All the stars have a relatively narrow component centred between +10 and~$+15\:{\rm km}\,{\rm s}^{-1}$. All the stars with spectroscopic distances similar to the cluster members have a broader component at higher velocities, with a peak in the $+40$ to~$+50\:{\rm km}\,{\rm s}^{-1}$ range. For all of them, this high-velocity component reaches a maximum of $\approx+62\:{\rm km}\,{\rm s}^{-1}$ close to the continuum, while at half maximum intensity has a value of $\approx+53\:{\rm km}\,{\rm s}^{-1}$. All the cluster members observed, plus HD~292163, HD~292164, and HD~292167, have very similar profiles, sharing all these characteristics (see Fig.~\ref{redhelium}). We can readily identify this high-velocity component as arising from material associated with Sh2-284, given that it has the same peak velocity and a similar width to the H$\alpha$ emission coming from the \ion{H}{ii} region. This is borne out by the fact that HD~292090, with a distance similar to the cluster, but rather far away from it, displays weaker absorption at these velocities.

Stars with lower distances lack this high-velocity component. The nearby stars HD~48619 and HD~292398 have a single component at positive velocities (see left panel of Fig.~\ref{NaD}), centred on $+12\:{\rm km}\,{\rm s}^{-1}$, which coincides with one of the peaks in the spectra of the stars at higher distances. This velocity corresponds to a distance of $\sim1.2$~kpc in the rotation curve, in good agreement with the spectroscopic distances of HD~48619 and HD~292398. The stars HD~48807 and HD~292392, which are located at intermediate distances, show absorption components centred around $+30\:{\rm km}\,{\rm s}^{-1}$. This value corresponds to $\sim3$~kpc in the rotation curve, again in very good agreement with their distances (see next section). 

Therefore the radial velocities of the cluster members and the interstellar material seen in their spectra place the whole cluster + \ion{H}{ii} region complex at $\sim5$~kpc. Naturally, deviations from circular rotation are possible. In this respect, the maser source G211.59+01.05 has a parallax distance of 4.4~kpc with rather small formal errors \citep{reid14}. This source is associated with IRAS~06501+0143, with a radial velocity $v_{\mathrm{LSR}}=+45.5\:{\rm km}\,{\rm s}^{-1}$ \citep{bronfman96}. It is only about $2\degr$ away from Sh2-284. Given this small on-the-sky separation and that it has exactly the same radial velocity, it seems sensible to assume that Sh2-284 is at the same distance.

Finally, in a companion paper (Lorenzo et al. 2015), we study the eclipsing binary GU~Mon, which is a cluster member (S13). Under the assumption of spherical stars, we derive a distance of $3.9\pm0.4$~kpc, which should be treated as a lower limit because of the extreme compactness of the system. Given the consistency of all these estimates, we assume a distance $4.5\pm0.3$~kpc for the cluster.

\subsection{The sHRD}

 To check the consistency of the analysis, we used the parameters derived for all the stars to build the spectroscopic  Hertzsprung--Russell diagram (sHRD). The sHRD is obtained from the HRD by replacing the luminosity ($L$) for the new variable  ${\mathscr L} := T_{\rm eff}^4/g$, i.e. the inverse of the flux-weighted gravity introduced by \citet{kudritzki03}. This means that the sHRD, unlike the classical HRD, can be drawn entirely using results of the atmosphere fits, without having to assume a distance for the objects analysed \citep{lankud14,castro14}. Moreover, the sHRD is independent of the effects of extinction.

\begin{figure}
   \centering
\resizebox{\columnwidth}{!}{\includegraphics[clip]{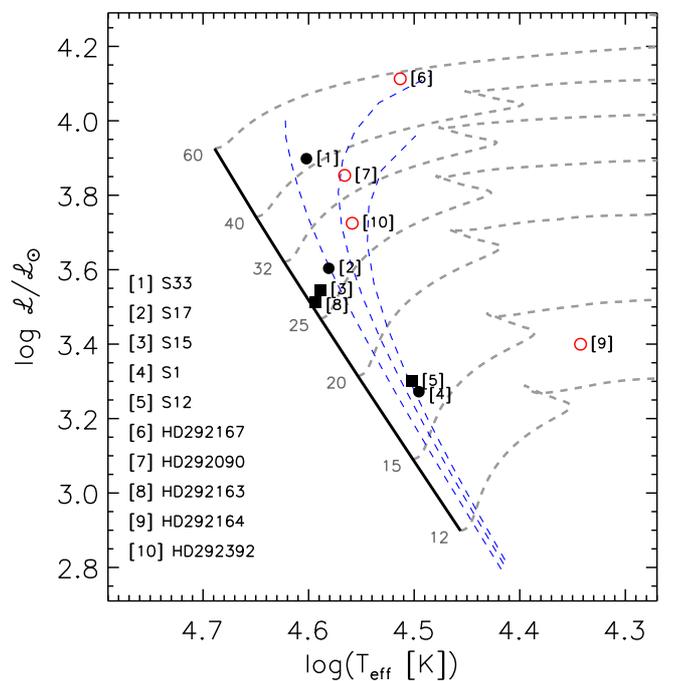}}
   \caption{Spectroscopic HR diagram for the stars analysed ({\bf top}). Squares mark stars identified as binaries, or suspected of binarity. Stars in red are not considered cluster members. Evolutionary tracks from \citet{ekstrom12} with $v_{\textrm{rot,} 0}=0$ and solar metallicities are shown as references. Isochrones from the same source correspond to $\log t = 6.3, 6.5$, and 6.6. 
              \label{fig:shrd}}
    \end{figure}

The position of the stars in the sHRD is shown in Fig.~\ref{fig:shrd}  We did not consider HD~292398, which is not really a high-mass star, or HD~48691 because its distance is much shorter than that of the cluster. HD~48807 is not shown, as this would imply changing the scale and it is certainly not a cluster member. Its position in the sHRD diagram suggests that HD~48807 is slightly brighter than a simple adoption of the \citet{turner80} calibration, though within the intrinsic dispersion expected for supergiants.

When using the classical HRD at the cluster distance, the loci of the stars with respect to the evolutionary tracks is the same, except for HD~292167. Even though its parameters are typical of stars of similar spectral type, its position in the sHRD places it close to the $60\,M_{\sun}$ track, whereas if we use the HRD (assuming the adopted distance to the cluster), it falls just above the $40\,M_{\sun}$ track, suggesting then an age slightly above 3~Myr. At this luminosity, it would need to be rather farther away than the cluster ($\sim6\:$kpc) to reconcile both diagrams. However, its spectroscopic distance and interstellar lines suggest that it is at the same distance as the cluster. This could be another case of the well-known discrepancy between spectroscopic and evolutionary masses \citep[see, e.g.][]{herrero07}, or could be due to difficulties with the spectroscopic analysis or undetected binarity. We note that the fit in Fig.~\ref{fitO} is slightly poorer than for the other stars (see \ion{He}{ii}~4686\AA), while its rotational velocity,  $v_{\textrm{rot}}=150\:\textrm{km}\,\textrm{s}^{-1}$, is very high for a supergiant \citep[see][for typical values]{ssimon14}.

\subsection{Abundances in Dolidze 25 and the Galactic Anticentre}

\citet{lennon90} conclude that Dolidze 25 is defficient in metals by approximately a factor of six with respect to the Sun. This conclusion was reached via a differential analysis of the spectra of S15, S17, and S22 compared to 10~Lac and $\tau$~Sco, and an LTE analysis of S22. As we have shown above, the spectral types of S15 and S17 are too early for a meaningful comparison to 10~Lac. These stars lack metallic lines because they are too hot to have any, except \ion{Si}{iv}~4089\AA. However, a re-analysis of data for S22 by \citet{rolleston94} still leads to a mean metallicity $-0.7\pm0.1$~dex below the solar value via comparison to other stars in the Anticentre region, almost all of which, however, have rather higher metallicities (not very different from those observed in the solar neighbourhood). In particular, they find the sub-abundance of Si to be extreme. Finally, a new analysis of the data by \citet{rolleston00}, this time using new LTE and non-LTE models, still finds abundances of O and Si between $-0.4$ and $-0.5$~dex lower than in stars of clusters at the same Galactocentric distance.

We have not derived abundances for S22 because its relatively high rotational velocity $v_{\mathrm{rot}}$ makes it a less-than-ideal target for these purposes. Our spectrum, with a resolution similar to that used by \citet{lennon90} but much higher S/R, does not allow an accurate derivation of stellar abundances. We have instead derived parameters and abundances from two other stars of similar spectral type, located in the outskirsts of the cluster (namely, S1 and S12), and then made sure that the spectrum of S22 is compatible with the abundances derived. We find a Si abundance $-0.3$~dex and an O abundance $-0.5$~dex lower than the typical abundances in the solar neightbourhood. This means that our results support the low metal content of Dolidze~25, even though it is not as extreme as originally assumed, and more in line with the values reported by \citet{rolleston00}. Part of the discrepancy with respect to the values found by \citet{lennon90} is simply due to the fact that the currently accepted values for the solar abundances are lower than assumed at the time of their work. Therefore, even though the actual abundances measured are not much higher, the difference with respect to the solar neighbourhood is smaller.

 The abundances that we have found for Do~25 are roughly compatible with what is expected from current values of the Galactic metallicity gradient. For a Galactocentric distance $r_{\textrm{G}}\approx12\:$kpc, the gradients found by \citet{rolleston00}, $-0.06$\:dex\,kpc$^{-1}$ for Si and $-0.07$\:dex\,kpc$^{-1}$ for O, imply subsolar abundances by $-0.24$~dex in Si and $-0.27$~dex in O. The abundance gradient for O based on optical studies of \ion{H}{ii} regions has a similar slope \citep{rudolph06}. Our values for Si are compatible within their errors, with the values for O still remaining too low (at almost 2$\sigma$ from their expected value). The lower abundance of oxygen with respect to silicon may be real, even if they are compatible within their respective errors, since the two stars analysed give almost identical values. However, we have to emphasise that at least S12 shows the signatures of a fainter companion in its spectrum, and therefore all abundances derived must be considered lower limits, with the effect of the companion on the abundances dependent on its spectral type. Since the companion is fainter, but still detectable, it is likely to have a spectral type earlier than B2. If so, the \ion{Si}{iii} and \ion{O}{ii} lines should be intrinsically much stronger than in the O9.7\,V primary. This results in a stronger contribution to the lines than expected from its relative flux. In spite of this, the lines of the companion are not very strong compared to those of the primary. Considering typical values for the absolute magnitude, the companion cannot contribute more than 30\% of the flux, and likely will be at the $\sim$20\% level. If so, the abundances derived may be actually increased by up to 20\% (i.e. increase the oxygen abundance from $-$0.5 to $-$0.4~dex), assuming that we have been able to exclude completely the contribution of the companion to the shape of the lines (if not, the increase will be slightly smaller).

On the other hand, the abundances that we find for HD\,292164, which is at least at the same distance as Do~25 and has a spectrum much more adequate for deriving accurate values, are only $-0.1$~dex lower than solar values, and thus compatible with them within the uncertainties. \citet{rolleston00} also find very mild abundance defficiencies for other clusters at similar Galactocentric radii, such as NGC~1893. In particular, we note that Sh2-285, located only $\sim$$2\fdg5$ away from Do~25, has essentially the same radial velocity $v_{{\rm CO}}=+45.3\pm1.1\:{\rm km}\,{\rm s}^{-1}$, $v_{{\rm H}_{\alpha}}=41.2\pm0.5\:{\rm km}\,{\rm s}^{-1}$ \citep{blitz82,fich90}. From the analysis of two early-B stars in this cluster, \citet{rolleston94} concluded that metallicity was only very slightly subsolar. However, given its radial velocity and its spectroscopic distance, this cluster is almost certainly at the same distance as Do~25.

This result, however, is not completely surprising, as there seems to be a degree of spread in the metallicities found at a given Galactocentric radius. For example \citet{lemasle13} derive abundance gradients for a large number of elements from an analysis of Cepheid variables. In particular, for Si, they find a very good agreement with the slope determined by \citet{rolleston00}. However, \citet{genovali14} study the spread of individual values about the average gradient, finding a steady increase in the abundance dispersion when moving into the outer disk. This effect becomes more evident beyond $r_{\textrm{G}}\approx14\:$kpc. From correlations between spatial position and abundances, \citet{genovali14} conclude that real variations exist on scales that range from the size of an OB association to a stellar complex in a spiral arm. Even the more accurate data of \citet{genovali15} present variations of a few tenths of a dex between stars at a given Galactocentric radius. In view of this, we can conclude that the abundances determined from early-type stars at $r_{\textrm{G}}\approx12\:$kpc are in agreement with currently accepted values of the metallicity gradient slope, with Do~25 representing an extreme of the intrinsic dispersion (in the sense of lower metal content), with Sh2-285 and HD~292164 representing the other.

\subsection{Properties and extent of the star-forming region}

A cursory look at Fig.~\ref{CMD} demonstrates the variability of reddening across the cluster. There is a significant number of members, in particular S12, S13, and S22, that display a moderate reddening, not far from $E(J-K_{\mathrm {S}})=0.36$, compatible with those of field stars at similar distances, both in the cluster area (e.g. HD~292164) or at some distance (e.g. HD~292090, HD~292392). In contrast, the B8\,Ib supergiant HD~48807, which is clearly foreground to the cluster, but $\ga 3$~kpc away from the Sun, suffers a reddening of only $E(J-K_{\mathrm {S}})=0.16$. 

On the other hand, many other cluster members display rather higher reddening, with a most typical value $E(J-K_{\mathrm {S}})\approx0.55$. Comparison of the values of $E(B-V)$ and $E(J-K_{\mathrm {S}})$ suggest that S9 and S33 have an infrared excess (though the optical photometry for S33 is of low quality, and it could be simply slightly more reddened than the other members). For all the other members with spectroscopy and indeed all other stars in the field without emission lines, $E(B-V)/E(J-K_{\mathrm {S}})=0.52\pm0.06$ (one standard deviation), in perfect agreement with the standard extinction law.

Na\"ively, one could say that stars near the cluster core (S15, S16, S17) are more reddened than those towards the periphery, but this is not necessarily so, as S33 is in the periphery of the cluster, and S1 and S9 are located on the \ion{H}{ii} shell. In any case, the definition of a typical (or average) reddening for the cluster does not seem to be granted, a situation not unexpected for a cluster immersed in nebulosity.

The cluster as studied by \citet{delgado10} lies close to the centre of the \ion{H}{ii} region. However, some of our targets lie on the inner rim of the \ion{H}{ii} shell, while our identification of two B1.5\,V stars close to IRAS~06413+0023 clearly shows that there are early-type stars beyond this rim. How far does the star-forming region extend?

As discussed above, the answer to this question involves dilucidating whether HD~292167 is a cluster member. This is the most luminous star in our sample, and almost certainly the most massive. Its radial velocity is somewhat lower than that of the cluster and \ion{H}{ii} region. A lower radial velocity, however, suggests a lower distance, which is not supported by the spectroscopic analysis. We must thus speculate that the star either has a peculiar velocity (and so may have been ejected from the cluster) or perhaps is an undetected binary (we only have one radial velocity measurement), a condition that would also explain its luminosity (which, taken at face value, places it beyond the cluster). Circumstancial evidence very strongly points to a physical connection. The optical and WISE images (Fig.~\ref{map}) clearly show the \ion{H}{ii} and dust emission extending to the north of the main shell. The infrared source IRAS~06425+0038, corresponding to the densest part of this Northern extension, is located just $3\arcmin$ away from HD~292167 (which thus seems to be its ionisation source) and has the same radial velocity as Sh2-284.

\begin{table}
 \centering
 \begin{minipage}{\columnwidth}
  \caption{IRAS sources in the immediate vicinity of Sh2-284 whose radial velocities suggest a physical association \citep[from][]{wouter89}.\label{iras}}
  \begin{tabular}{lcccc}
  \hline
\hline
 \noalign{\smallskip}
Name   & $l$&$b$ & $v_{{\rm LSR}}$ & $\Delta v$    \\
 \noalign{\smallskip}
 \hline
 \noalign{\smallskip}
06422+0023& 211\fdg87&$-$02\fdg32&+44.11$\pm0.02$&2.94\\
06423+0006& 212\fdg13&$-$01\fdg43&+46.83$\pm0.05$&2.67\\
06425+0038& 211\fdg68&$-$01\fdg13&+44.10$\pm0.10$&4.00\\
06437+0009$^{a}$& 212\fdg25&$-01\fdg10$&+42.01$\pm0.02$&4.55\\
06445+0009& 212\fdg35&$-$00\fdg92&+44.32$\pm0.03$&3.96\\
06446+0029& 212\fdg06&$-$00\fdg74&+44.30$\pm0.10$&4.60\\
06454+0020& 212\fdg29&$-$00\fdg62&+46.41$\pm0.04$&2.25\\
06458+0040& 212\fdg03&$-$00\fdg40&+43.73$\pm0.06$&2.07\\
 \noalign{\smallskip}
\hline
\end{tabular}
\begin{list}{}{}
\item[]$^{a}$ There is a much weaker component at $v_{{\rm LSR}}=+27.3\:\mathrm{km}\,{\rm s}^{-1}$.
\end{list}
\end{minipage}
\end{table}

As a matter of fact, there are many IRAS sources in the area identified as compact \ion{H}{ii} regions in the survey of \citet{wouter89} that share very similar radial velocities (see Table~\ref{iras}). These sources extend over almost two degrees to the south of the Galactic Plane. One of them is IRAS~06446+0029, about $35\arcmin$ away from the cluster, which harbours the emission-line star SS~62, discussed above. Slightly more distant \ion{H}{ii} regions, such as GB6~B0642+0111 ($\sim50\arcmin$ north) or NGC~2282 ($\sim70\arcmin$ NE) have distinctly lower radial velocities \citep{blitz82}, and are therefore very likely closer to the Sun than Sh2-284. However, towards the east, the structure seems to extend farther away. IRAS~06454+0020 has a very similar velocity $v_{{\rm LSR}}=+46.4\:\mathrm{km}\,{\rm s}^{-1}$ (Table~\ref{iras}). A few arcminutes to the east lies the open cluster Bochum~2, discovered by  \citet{moffat75}, who assigned it a distance compatible with that of Do~25. The two brightest stars are spectroscopic binaries with spectral types around O9 \citep{munarito99}. From their orbital solutions, \citet{munarito99} derive a cluster $v_{\textrm{hel}}=+68\pm3\:\mathrm{km}\,{\rm s}^{-1}$, equivalent to $v_{{\rm LSR}}\approx+51\:\mathrm{km}\,{\rm s}^{-1}$. \citet{turbide} quote $v_{{\rm LSR}}=+49\pm7\:\mathrm{km}\,{\rm s}^{-1}$ after \citet{jackson79}. Both values are compatible among themselves and not very different from the velocity for Sh2-284\footnote{The uncertainty quoted by \citet{munarito99} is likely to be an underestimation, as it is based on only three measurements of radial velocities, two of them systemic velocities of early-type spectroscopic binaries.}. Therefore it seems clear that Sh2-284 is part of a rather extended region of star formation covering close to one square degree in the sky.

\section{Conclusions}

We have carried out a comprehensive spectroscopic survey of stars in and around the open cluster Dolidze~25, proposed to have a very low metallicity, and combined our results with existing photometry to derive stellar and cluster parameters. From our analysis, we find that:

   \begin{enumerate}
 \item Dolidze~25 has a low metallicity, with values $-0.3$~dex below solar for Si and $-0.5$~dex below solar for O. Even though these values are not as low as those found by \citet{lennon90}, Do~25 is confirmed as the star-forming site with a lowest metallicity known in the Galaxy. Unfortunately, these values cannot be considered as secure as those obtained for other regions \citep[i.e.][]{ssimon10}, as the number of stars with adequate parameters in the cluster is small. S12, analysed here, shows evidence of a fainter companion, while S22, used by previous authors, is fainter (meaning that we had to settle for a lower-resolution spectrum) and presents a moderately high projected rotational velocity, contributing to make its analysis complicated. Only S1 seems to be free of complications, and it is very far away from the central region of the cluster. Even then, close inspection of its spectrum does not entirely leave out the possibility of a faint companion, although its radial velocity is consistent with the nebular values. Perhaps some of the likely members with spectral types close to B1\,V (i.e. S16, S18) could also be used to derive abundances, but with $B>14$~mag, these objects require much larger telescopes. 
\item This low metallicity is not inconsistent with modern values for the slope of the metallicity gradient, as some dispersion at a given Galactocentric radius seems to exist. In fact, several other clusters at the same Galactocentric radius were found to have higher metallicities \citep{rolleston00}. The only narrow-lined star analysed in this work, HD~292164, which is at approximately the same distance as the cluster, has a metallicity only very slightly subsolar (and barely compatible with solar metallicity within the errors).
\item The ionising photons powering the \ion{H}{ii} region Sh2-284 are provided by the O6\,V star S33 and the O7\,V stars S15 and S17. None of these objects shows evidence of any evolution, implying an upper limit of $\sim3$~Myr for the cluster age, compatible with age estimates of the PMS population. Given the absence of any evolved stars, an accurate age cannot be given. The O8.5\,Ib supergiant HD~292167 is located to the north of the cluster, and is likely to be physically connected, even though the spectroscopic analysis suggests a slightly higher distance. Its position in the tracks is also consistent with an age $\sim3$~Myr.
\item Several lines of evidence suggest that the cluster is located at a distance between 4 and 5~kpc. We adopt the trigonometric parallax distance to the nearby \ion{H}{ii} region IRAS~06501+0143, and settle for a distance $d=4.5\pm0.3\:$~kpc.
      \item There is no evidence for the evolved population suggested by \citet{delgado10} in the central concentration generally identified as Do~25. Stars 15 and 17, which they assume to be B1 giants, are O7 dwarfs. There are, however, many OB stars around the \ion{H}{ii} shell suggestive of different ages. In particular, HD~292164 (which is indeed a B1.5\,III giant) is necessarily rather older.
\item At least two emission-line stars located within the \ion{H}{ii} shell are excellent candidates to very early Herbig Be stars. Both objects are associated with bright mid-IR luminosity. The presence of these very young stars confirms that star formation is still taking place close to the photo-illuminated rim, as suggested by \citet{puga09} from the analysis of \textit{Spitzer} sources.
\item Dolidze~25 seems to be part of a rather large star-forming complex, of which Sh2-284 is the main component. As suggested by \citet{puga09}, other smaller \ion{H}{ii} regions, such as IRAS~06439$-$0000 (Cluster~3 in \citealt{puga09}), IRAS~06446+0029 or IRAS~06454+0020 are part of this region. The presence of masers in this latter source (GAL~212.25$-$01.10) offers an excellent opportunity to measure a more accurate distance to the region.

   \end{enumerate}

\begin{acknowledgements}

Partially based on observations made with the Nordic Optical Telescope, operated by the Nordic Optical Telescope Scientific Association, and the Mercator Telescope, operated by the Flemish Community, both at the Observatorio del Roque de los Muchachos (La Palma, Spain)
of the Instituto de Astrof\'isica de Canarias. The INT and WHT are operated on the  island of La Palma by the Isaac Newton Group in the Spanish Observatorio del Roque de Los Muchachos of the Instituto de Astrof\'{\i}sica de Canarias. Some of the WHT and NOT observations were taken as part of their respective service programmes, and we would like to thank the staff astronomers for their diligence.

 This research is partially supported by the Spanish Ministerio de Econom\'{\i}a y Competitividad under 
grants AYA2012-39364-C02-01/02, and the European Union. This research has made use of the Simbad, Vizier 
and Aladin services developed at the Centre de Donn\'ees Astronomiques de Strasbourg, France. This research has made use of the WEBDA database, operated at the Department of Theoretical Physics and Astrophysics of the Masaryk University.

This paper makes use of data obtained from the Isaac Newton Group Archive which is maintained as part of the CASU Astronomical Data Centre at the Institute of Astronomy, Cambridge.
\end{acknowledgements}

%\begin{thebibliography}{}

% \bibitem[1966]{baker} Baker, N. 1966,
%      in Stellar Evolution,
%      ed.\ R. F. Stein,\& A. G. W. Cameron
%      (Plenum, New York) 333

%\end{thebibliography}
\bibliographystyle{aa}
\bibliography{clusters}

\end{document}